\documentclass{article}

\usepackage{threeparttable}
\usepackage{microtype}
\usepackage{graphicx}
\usepackage{subcaption}
\usepackage{makecell}
\usepackage{booktabs} 
\usepackage{tabularx}
\usepackage{multirow} 
 
\usepackage{hyperref}

\usepackage[preprint]{icml2026}

\usepackage{amsmath}
\usepackage{amssymb}
\usepackage{mathtools}
\usepackage{amsthm}

\usepackage[capitalize,noabbrev]{cleveref}

\theoremstyle{plain}

\theoremstyle{definition}

\theoremstyle{remark}

\usepackage[textsize=tiny]{todonotes}

\icmltitlerunning{Synthesizing Epileptic Seizures: Gaussian Processes for EEG Generation}

\begin{document}

\twocolumn[
  \icmltitle{Synthesizing Epileptic Seizures: Gaussian Processes for EEG Generation}



  \icmlsetsymbol{equal}{*}

  \begin{icmlauthorlist}
    \icmlauthor{Nina Moutonnet}{xxx}
    \icmlauthor{Joshua Corneck}{yyy}
    \icmlauthor{Felipe Tobar}{yyy}
    \icmlauthor{Danilo Mandic}{zzz}
  \end{icmlauthorlist}
  \icmlaffiliation{xxx}{Department of Computing, Imperial College London, UK}
  \icmlaffiliation{yyy}{Department of Mathematics, Imperial College London, UK}
  \icmlaffiliation{zzz}{Department of Electrical and Electronic Engineering, Imperial College London, UK}

  \icmlcorrespondingauthor{Nina Moutonnet}{nina.moutonnet18@imperial.ac.uk}

  \icmlkeywords{Machine Learning, ICML}

  \vskip 0.3in
]



\printAffiliationsAndNotice{}  

\begin{abstract}

Reliable seizure detection from electroencephalography (EEG) time series is a high-priority clinical goal, yet the acquisition cost and scarcity of labeled EEG data limit the performance of machine learning methods. This challenge is exacerbated by the long-range, high-dimensional, and non-stationary nature of epileptic EEG recordings, which makes realistic data generation particularly difficult. In this work, we revisit Gaussian processes as a principled and interpretable foundation for modeling EEG dynamics, and propose a novel hierarchical framework, \textit{GP-EEG}, for generating synthetic epileptic EEG recordings. At its core, our approach decomposes EEG signals into temporal segments modeled via Gaussian process regression, and integrates a domain-adaptation variational autoencoder. We validate the proposed method on two real-world, open-source epileptic EEG datasets. The synthetic EEG recordings generated by our model match real-world epileptic EEG both quantitatively and qualitatively, and can be used to augment training sets.

\end{abstract}

\section{Introduction}

Electroencephalography (EEG) is a non-invasive neuroimaging modality that is used to record cerebral electrical activity from scalp electrodes \cite{moutonnet2024clinical}. It is widely used for diagnosis and monitoring in neurology, particularly in epilepsy and sleep medicine \cite{attarian2012atlas, hirsch2011atlas, rasheed2021, dash2025, li2025}. Although machine learning (ML) methods for EEG have advanced rapidly \cite{gui2024, afzhal2024, covert2019, ahmedt2020}, performance on specialized clinical tasks, such as seizure detection, is often limited by scarce and heterogeneous labeled data. In contrast to more regularly observed EEG phenomena (e.g., sleep staging or steady-state visually evoked potentials), epileptic seizures are rare and unpredictable, and both interictal and ictal signatures vary substantially across patients \cite{kamrud2021, vos2025}. Ethical, logistical, and clinical constraints further restrict large-scale, systematic acquisition and sharing of high-quality recordings \cite{ney2024, boudewyn2023, arora2025}. These challenges motivate high-fidelity synthetic EEG generation as a practical route to augment training sets and evaluate algorithms for clinically important but infrequent events \cite{rujas2025, murtaza2023, pantanowitz2024, juwara2024}.

Most synthetic EEG generation methods rely on deep generative models, including generative adversarial networks  \citep[GANs;][]{hazra2020synsiggan, haradal2018biosignal, hartmann2018eeg, pascual2020epilepsygan}, variational autoencoders \citep[VAEs;][]{aznan2019simulating}, and diffusion models \cite{aristimunha2023synthetic, torma2025generative, vetter2024generating}. While these approaches can reproduce both realistic waveforms and spectral characteristics, prior work typically targets short segments and/or a limited subset of EEG channels, relies on windowed training that does not scale to continuous recordings, and is difficult to interpret or to steer toward clinically meaningful scenarios \cite{seyfi2022generating, hazra2020synsiggan, vetter2024generating}. Clinically realistic EEG synthesis instead demands: (i) long-horizon generation that captures non-stationary evolution over time, and (ii) preservation of multivariate spatial structure across the full electrode montage \cite{chaddad2023electroencephalography, baier2012importance, zhong2022temporal}.

This paper addresses long-horizon synthesis of multi-channel EEG recordings. The problem is challenging because EEG is strongly non-stationary, its amplitude, phase, and spectral structure evolve over time in complex ways, and it exhibits temporal dependencies spanning multiple scales, including long-range correlations \cite{baier2012importance, zhong2022temporal, hong2025samba}. In addition, clinically recorded EEG is inherently multivariate (e.g., standard recordings typically use 19 scalp electrodes), requiring preservation of structured inter-channel dependencies in the generated signals \cite{vetter2024generating, hong2025samba}. Finally, generating sequences with thousands of samples per channel at clinical sampling rates is computationally demanding for end-to-end neural generators and is further complicated by practical considerations such as interpretability, as well as limited access to training data \cite{lyu19deep, vetter2024generating}.

To address these issues, we propose GP-EEG, an interpretable pipeline for generating long-horizon, multivariate, patient-specific EEG. We first simultaneously compress all of the recordings of a patient into a low-dimensional representation via singular value decomposition (SVD), yielding \textit{temporal scores} and corresponding \textit{spatial loadings}. Temporal scores are segmented into quasi-stationary regimes using offline changepoint (CP) detection, and a Gaussian process (GP) model is fit within each regime to capture local temporal structure with uncertainty quantification. For generation, we sample a CP sequence from a Poisson process trained on real EEG data and model the state transitions between GP kernel configurations (defined by their optimized hyperparameters) using a discrete Markov chain, thus generating temporal scores from a sequence of regime-specific GP samples. We then reconstruct synthetic multi-channel recordings by projecting the generated temporal scores through the patient-specific SVD spatial loadings, preserving realistic inter-channel correlation. Finally, we apply a convolutional long short-term memory VAE (Conv-LSTM VAE) as a domain adaptation step to better match characteristic EEG spectral content and waveform morphology, while retaining the prescribed long-range temporal evolution and spatial dependencies.

The proposed methodology is evaluated on two public EEG datasets, CHB-MIT \citep{shoeb2009application} and Siena \citep{detti2020siena}, and show that it generates realistic long-horizon, multi-channel recordings that preserve temporal statistics, inter-channel relationships, and clinically relevant signal characteristics.

Our contributions are as follows:
\begin{enumerate}
\item We introduce an interpretable framework, GP-EEG, for patient-specific generation of long-horizon, multivariate EEG recordings (thousands of time points per channel at clinical sampling rates), combining low-dimensional compression with explicit regime-based temporal modeling.
\item We preserve multivariate spatial structure by generating temporal scores and reconstructing full-montage recordings using patient-specific spatial loadings, maintaining realistic inter-channel dependencies.
\item We couple changepoint-based segmentation with regime-wise GP modeling and an explicit generative model of regime dynamics, enabling tractable long-horizon synthesis with stage-wise interpretability and control.
\item The code will be made available after anonymization is lifted.
\end{enumerate}

The remainder of this paper is organised as follows: Section \ref{sec:related_lit} provides a review of existing research on synthetic EEG data generation. Section \ref{sec:methodology} presents our methodology. Section \ref{sec:results} reports experimental results. Finally, Section \ref{sec:conclusion} discusses the benefits and limitations of our method, and outlines potential directions for future work.

\section{Related literature} \label{sec:related_lit}

\begin{figure*}[h!]
    \centering
    \includegraphics[width=0.95\textwidth]{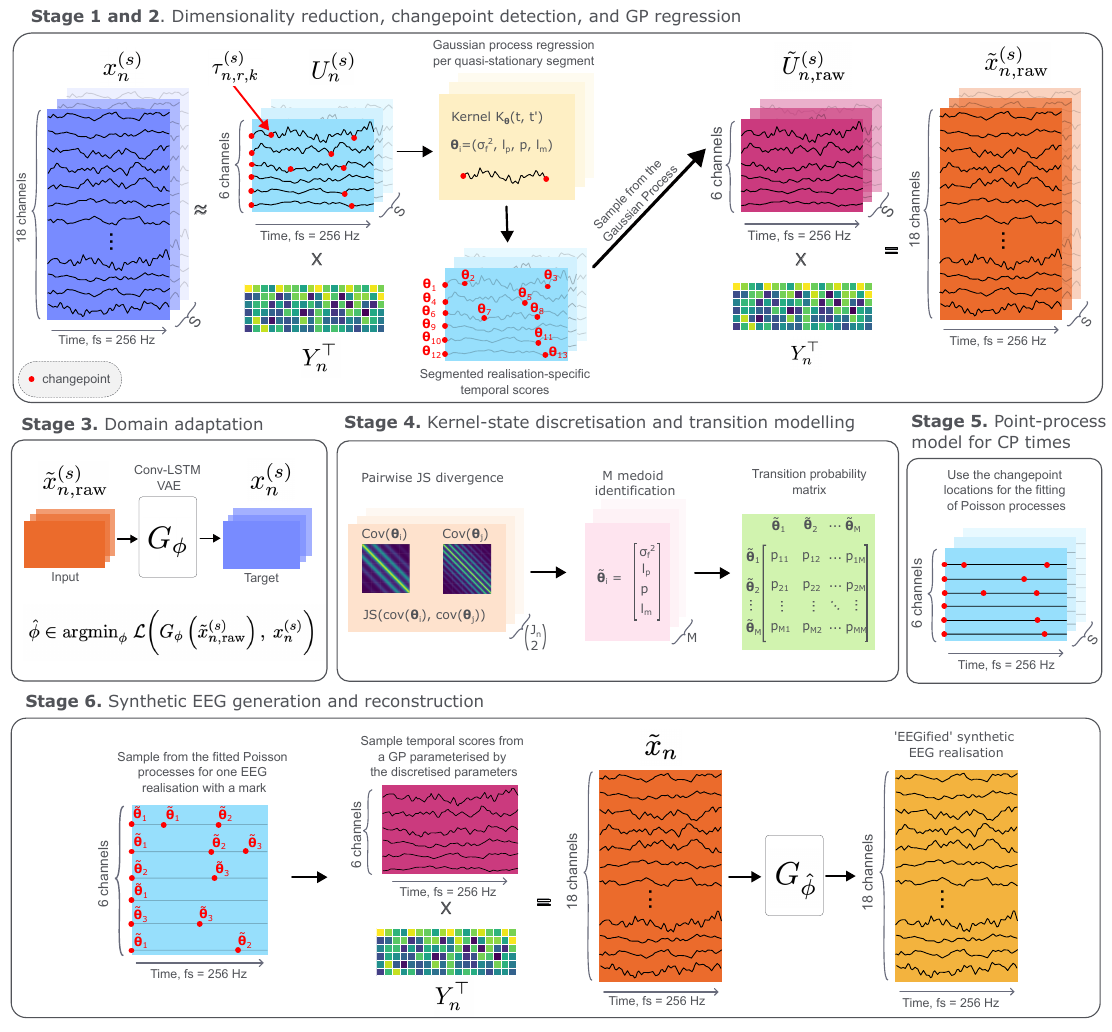}
    \caption{Overview of GP-EEG, the proposed multi-stage framework for patient-specific synthetic EEG generation, described in Section \ref{sec:full_pipeline}. When time series are colored, dark blue corresponds to real seizures, dark orange to the output of Stage 2 and 6 prior to `EEGification', and yellow to that of Stage 6 following `EEGification'.}
    \label{fig:summary_diagram}
\end{figure*}

Modeling long, multivariate time series presents unique challenges. Early approaches relied on recurrent neural networks \cite{rumelhart1985learning}, but their applicability to long sequences was limited by vanishing and exploding gradients \cite{pascanu2013difficulty}. LSTM networks partially alleviated these issues through gated recurrence, enabling better modeling of long-range dependencies \cite{hochreiter1997long} and are commonly used in time series GANs \cite{harada2019biosignal}. Transformers \cite{vaswani2017attention} removed recurrence entirely, allowing parallel processing via self-attention and enabling advances in time series forecasting, classification and imputation \cite{wang2024medformer, yildiz2022multivariate}. Recent state-space models (SSMs), including linear SSMs and neural variants such as S4 and Mamba \cite{gu2021efficiently, gu2024mamba}, can capture long-range dependencies. However, both transformers and SSMs remain largely unexplored in generative contexts, especially for high-dimensional, multichannel signals where training data is scarce \cite{naiman2024utilizing}.

Deep generative models, including GANs, VAEs, and more recently diffusion models, have been applied to sequential data. TimeGAN \cite{yoon2019time} synthesizes sequences by jointly training a generator and discriminator while minimizing an autoregressive, unsupervised loss in the embedding space. Despite these advances, most GAN- and VAE-based models struggle to scale to long-horizon, multivariate generation. Diffusion models for time series generation can generate high-quality sequences while capturing temporal and spectral structure, as shown in \citet{vetter2024generating}. Building on the success of diffusion models in image generation, \citet{naiman2024utilizing} recently proposed transforming time series into images using invertible transforms (short-time Fourier transform and delay embeddings), enabling diffusion-based generation of short, long and ultra-long sequences. However, score-based generative modeling requires accurate estimation of high-dimensional score functions, which typically needs large training datasets, particularly when modeling complex data distributions \cite{dhariwal2021diffusion}.

Several works have applied generative models specifically to EEG. \citet{hazra2020synsiggan} proposed SynSigGAN, using a bidirectional grid LSTM generator and CNN discriminator, but the generated signals were univariate and extremely short (19 samples at 512 Hz, i.e., 37.1 ms). \citet{hartmann2018eeg} developed a CNN-CNN Wasserstein GAN to generate 3000-sample EEG segments from channel FCC4h, preserving spectral characteristics for left-hand movement. \citet{haradal2018biosignal} proposed an LSTM-LSTM GAN for generating univariate epileptic EEG segments up to 178 samples, and \citet{pascual2020epilepsygan} developed a CNN-CNN GAN for ictal EEG segments. COSCI-GAN \cite{seyfi2022generating} combines channel-specific LSTM generators and discriminators with a central multilayer perceptron (MLP) discriminator to enforce inter-channel correlations, generating 1200-samples EEG segments across 5 channels (fs = 128 Hz). While these methods improve short-window generation and inter-channel coherence, sequences remain far shorter than clinically recorded EEG. Diffusion-based EEG models \cite{aristimunha2023synthetic, vetter2024generating} produced realistic output, but are also generally applied to short windows and/or modest channel counts and require large training datasets.

\section{Methodology} \label{sec:methodology}

\subsection{Problem statement}

\textbf{General problem statement. \ }
We consider the problem of generating long and ultra-long, non-stationary, high-dimensional time-series (TS) data. We observe $S$ samples $\{x^{(s)}\}_{s=1}^S$ with
$x^{(s)} \in \mathbb{R}^{T^{(s)} \times C}$, where $T^{(s)}$ is the number of time points and $C$ the number of channels, and both $T^{(s)}$ and $C$ may be large. Each sample is assumed to be drawn from the same unknown distribution $p(x)$. Our goal is to generate synthetic samples $\tilde{x}\in\mathbb{R}^{T' \times C}$ whose distribution is close to $p(x)$, so that $\tilde{x}$ is statistically similar to real observations.

\textbf{EEG-specific problem statement. \ }
We observe EEG time-series from $N$ patients, each recorded on $C=18$ bipolar channels (corresponding to 19 electrodes), and we treat patients independently. Patient $n$ has $S_n$ recordings, denoted $\{x_n^{(s)}\}_{s=1}^{S_n}$ with $x_n^{(s)}\in\mathbb{R}^{T_n^{(s)}\times C}$, which we regard as draws from an unknown patient-specific distribution, $p_n(x)$. We aim to generate synthetic recordings $\tilde{x}_n\in\mathbb{R}^{T'\times C}$ that resemble samples from $p_n(x)$ both in terms of quantitative statistics and qualitative signal characteristics.

\subsection{Full synthesis pipeline} \label{sec:full_pipeline}

We now present our end-to-end pipeline that takes as input a collection of real EEG seizures, and produces as output a synthetic EEG seizure time series. The full pipeline is a six-stage framework (Figure~\ref{fig:summary_diagram}) that combines interpretable regime segmentation with scalable long-horizon generation. We describe these stages for a fixed patient $n$.

\paragraph{Stage 1: Dimensionality reduction.} Represent each multichannel recording as a matrix. Concatenate samples along the time axis to form a single matrix
\[
x_n \;=\; \mathrm{vstack}\!\left(x_n^{(1)},\dots,x_n^{(S_n)}\right)\in\mathbb{R}^{T_n\times C},
\]
where $T_n=\sum_{s=1}^{S_n}T_n^{(s)}$. Assume $x_n$ is column-centred. Apply singular value decomposition (SVD) to obtain a low-dimensional representation that captures the dominant shared structure across channels. Let the rank-$d$ truncated SVD of $x_n$ be $x_n \approx U_n D_n V_n^\top$ with $U_n\in\mathbb{R}^{T_n\times d},\;
D_n\in\mathbb{R}^{d\times d},\;
V_n\in\mathbb{R}^{C\times d}.$
We define the \emph{temporal scores} as $U_n$ and the \emph{spatial loadings} as $Y_n = V_nD_n\in\mathbb{R}^{C\times d}$. Writing the block of rows of $U_n$ corresponding to sample $s$ as
$U_n^{(s)}\in\mathbb{R}^{T_n^{(s)}\times d}$, we have
\[
x_n^{(s)} \approx U_n^{(s)} Y_n^\top.
\]
Further details are provided in Appendix \ref{app:dimred_svd_pca} and in Figure \ref{fig:appendix_svd}.

\paragraph{Stage 2: Changepoint detection and GP regression.} Offline changepoint (CP) detection is performed on the $d$-dimensional temporal scores to partition the recordings into quasi-stationary (QS) regimes. For the $s$-th sample, let $u_{n,r}^{(s)} \in \mathbb{R}^{T_n^{(s)}}$ denote the $r$-th column of $U_n^{(s)}$. Partition $u_{n,r}^{(s)}$ into $K_{n,r}^{(s)}+1$ QS segments. In our implementation, we combine evidence from the  Kwiatkowski-Phillips-Schmidt-Shin \citep[KPSS;][]{kwiatkowski1992testing} and Augmented Dickey Fuller \citep[ADF;][]{dickey1979distribution} stationarity tests to identify the QS boundaries (details in Appendix \ref{app_sec:CP_detection}). Let
\begin{align}
0=\tau_{n,r,0}^{(s)} &< \tau_{n,r,1}^{(s)} < \cdots \notag \\
&< \tau_{n,r,K_{n,r}^{(s)}}^{(s)} < \tau_{n,r,K_{n,r}^{(s)}+1}^{(s)} = T_n^{(s)}, \label{eqn:CP_times}
\end{align}
denote the resulting identified CPs, yielding $K_{n,r}^{(s)}+1$ segments. Within each regime, a GP regression model is fit to the temporal scores, using a fixed kernel family but regime-specific hyperparameters, thereby capturing local temporal structure with explicit uncertainty quantification. 
We use the quasi-periodic kernel
\begin{align}
    &k_{\theta}(t,t') = \sigma_f^2 \exp\left(-2\sin^2\left(\pi|t - t'|/p\right)/\ell_p^2\right)\notag\\
    &\times((1 + \sqrt{3}|t - t'|/\ell_m) \times \exp(-\sqrt{3}|t- t'|/\ell_m))\label{eq:kernel},
\end{align}
with $\theta=(\sigma_f^2,\ell_p,p,\ell_m)$. The kernel choice can be adapted to the data generation task at hand. A Mat\'ern-3/2 and a periodic kernel are combined. This choice follows from the need to model local periodicity and roughness of the EEG signal simultaneously, as described in Appendix \ref{app:choice_of_kernel}. The hyperparameter fitting procedure is described in Appendix \ref{app:hyperparam_grid_search}. For each component $r\in\{1,\dots,d\}$, the fitted kernel hyperparameter vector $\theta_{n,r,k}^{(s)}$ corresponds to the QS segment
\[
\mathcal{I}_{n,r,k}^{(s)} := \big(\tau_{n,r,k}^{(s)},\tau_{n,r,k+1}^{(s)}\big], \qquad k=0,\dots,K_{n,r}^{(s)}.
\]
For the $k$-th QS segment of component $r$, a synthetic segment
$\tilde{u}_{n,r,k}^{(s)}(\cdot)$ is generated by sampling from a GP with kernel \eqref{eq:kernel} and hyperparameter $\theta_{n,r,k}^{(s)}$:
\[
\tilde{u}_{n,r,k}^{(s)} \;\sim\; \mathcal{GP}\!\left(0,\; k_{\theta_{n,r,k}^{(s)}}(\cdot,\cdot)\right)
\quad \text{restricted to } \mathcal{I}_{n,r,k}^{(s)}.
\]
Sampling is performed independently across segments and concatenated to form a full-length synthetic temporal score for component $r$:
\[
\tilde{u}_{n,r}^{(s)}(t) \;=\; \sum_{k=0}^{K_{n,r}^{(s)}} \tilde{u}_{n,r,k}^{(s)}(t)\,\boldsymbol{1}\{t\in \mathcal{I}_{n,r,k}^{(s)}\},
\]
where $\boldsymbol{1}\{\cdot\}$ denotes the indicator function. These are then concatenated by stacking across $r$ to form the synthetic temporal-score matrix
\[
\tilde{U}_{n,\mathrm{raw}}^{(s)} := \big[\tilde{u}_{n,1}^{(s)},\dots,\tilde{u}_{n,d}^{(s)}\big]\in\mathbb{R}^{T_n^{(s)}\times d}.
\] 
When stacking, continuity of the trajectory is enforced by initializing QS segment $k+1$ with the endpoint of segment $k$. $\tilde{U}_{n,\mathrm{raw}}^{(s)}$ is projected back onto the 18-dimensional channel space using the patient-specific SVD spatial loadings from Stage~1:
\[
\tilde{x}_{n,\mathrm{raw}}^{(s)} \;=\; \tilde{U}_{n,\mathrm{raw}}^{(s)} Y_n^\top \;\in\;\mathbb{R}^{T_n^{(s)}\times C}.
\] 

There is now a one-to-one correspondence between a GP surrogate and the original data, with pairs $(\tilde{x}_{n,\mathrm{raw}}^{(s)},\; x_n^{(s)})$.

\paragraph{Stage 3: Domain adaptation.} The surrogate-original data pairs from Stage 2 are used as training data for a domain adaptation Conv-LSTM VAE, whose architecture is detailed in Table \ref{tab:vae_architecture}. This `EEGifies' the surrogate data to better match characteristic EEG spectral structure and waveform patterns, which may have been lost due to SVD compression and the inductive bias introduced by the GP with the chosen kernel, while preserving the generated long-range temporal and spatial dependencies.

Let $G_\phi:\mathbb{R}^{T\times C}\to\mathbb{R}^{T\times C}$ denote the Conv-LSTM VAE mapping with parameters $\phi$. We train $G_\phi$ as a domain-adaptation model using pairs $(\tilde{x}_{n,\mathrm{raw}}^{(s)},\; x_n^{(s)})$, where $\tilde{x}_{n,\mathrm{raw}}^{(s)}$ is treated as the input, and the original recording $x_n^{(s)}$ as the reconstruction target. We fit $\hat{\phi}$ by minimizing the standard VAE objective in a patient-specific manner. See Appendix \ref{app:vae_training} for training details.

\begin{table}[h]
\centering
\caption{Conv-LSTM VAE architecture for domain adaptation.}
\label{tab:vae_architecture}
\small
\begin{tabular}{@{}lll@{}}
\toprule
\textbf{Stage} & \textbf{Layer} & \textbf{Output} \\
\midrule
\multirow{3}{*}{Encoder} 
& Conv1d-BN-LReLU (18→64) & (B, 64, 512) \\
& Conv1d-BN-LReLU (64→128) & (B, 128, 256) \\
& Conv1d-BN-LReLU (128→256) & (B, 256, 128) \\
\midrule
\multirow{2}{*}{Latent}
& Conv1d (×2) + Pool & (B, 256) \\
& $z \sim \mathcal{N}(\mu, \sigma^2)$ & (B, 256) \\
\midrule
\multirow{2}{*}{Decoder}
& LSTM (×18) & (B, 1, 256) \\
& Linear (×18) & (B, 18, 1024) \\
\midrule
Output & Residual ($\lambda=0.3$) & (B, 18, 1024) \\
\bottomrule
\end{tabular}
\end{table}

\paragraph{Stage 4: Kernel-state discretisation and transition modelling.} The set of GP kernel hyperparameters learned across regimes is discretised into a finite collection of \emph{kernel states}. We then estimate a Markov transition matrix over these states, which enables tractable long-horizon sampling by reusing a learned library of regime types and their transition dynamics.

Stage~2 produces one hyperparameter vector per QS segment. Fitted hyperparameters across all $(s,r,k)$ triplets are collected into $\{\theta_{n,j}\}_{j=1}^{J_n}$, where 
\[
J_n=\sum_{s=1}^{S_n}\sum_{r=1}^{d}(K_{n,r}^{(s)}+1),
\]
is the total number of parameters for patient $n$. Here, each $j$ indexes an $(s,r,k)$ triplet and thus a $\theta_{n,r,k}^{(s)}$ vector. To enable tractable long-horizon generation, this continuous set is discretized into $M_n$ representative \emph{kernel states} $\{\bar{\theta}_n^{m}\}_{m=1}^{M_n}$ (here $M_n=50$). To do this, fix a time grid $\{t_g\}_{g=1}^{G}$ of 2560 evenly-spaced samples on [0,10] and for each $\theta_{n,j}$ form the covariance matrix
\[
\Sigma(\theta_{n,j}) \in \mathbb{R}^{G\times G},
\qquad
\Sigma(\theta_{n,j})_{gg'} = k_{\theta_{n,j}}(t_g,t_{g'}).
\]
A pairwise dissimilarity matrix between the Gaussian measures
$\mathcal{N}(0,\Sigma(\theta_{n,j}))$ is computed (in our implementation using the Jensen--Shannon divergence; see figures in Appendix \ref{app_sec:discretisation}), and agglomerative clustering (average linkage) is applied to obtain $M_n$ clusters. The medoid parameter for the $m$-th cluster defines $\bar{\theta}_n^{m}$.

Each QS segment is assigned a state label $z\in\{1,\dots,M_n\}$ via its cluster membership. For each fixed pair $(s,r)$, the QS segmentation induces an ordered state sequence over time; an empirical transition matrix $P_n$ is computed by counting transitions between consecutive segments and row-normalizing.
The $(\{\bar{\theta}_n^{m}\},P_n)$ pair defines the discrete kernel-state dynamics used in the generation stage.

\paragraph{Stage 5: Point-process model for CP times.} The timing of regime changes are modeled by fitting an \emph{inhomogeneous} Poisson process to the CP locations. This yields a time-varying intensity function that captures how the expected rate of CPs evolves over the recording, and provides a principled distribution from which to sample synthetic CP times for generation.

For a sample $s\in\{1,\dots,S_n\}$ and a component index $r\in\{1,\dots,d\}$, the CP boundaries from Stage 2 are treated as event times of an inhomogeneous Poisson process on $[0,T_n^{(s)})$ with intensity function $\lambda_{n,r}^{(s)}(t)\ge 0$. $\lambda_{n,r}^{(s)}(t)$ is estimated non-parametrically by applying a Gaussian kernel density estimator to the observed event times with a bandwidth of 0.5. 

\paragraph{Stage 6: Synthetic generation and reconstruction.} A sample $s$ and a choice of sample length $T' > 0$ is made. For each $r \in \{1,\dots,d\}$, synthetic CP times are generated by sampling from an inhomogeneous Poisson process on $[0,T')$ with intensity $\lambda_{n,r}^{(s)}(t)$, as fitted in Stage 5. For channel $r$, this gives synthetic change points $\{\{\tilde{\tau}_{r,k}\}_{k=1}^{\tilde{K}_r}\}_{r=1}^d$. Each CP $\tilde{\tau}_{r,k}$ is assigned a parameter by sampling a sequence of length $\tilde{K}_r$ from $P_n$. Repeat Stage 2 with these parameters and CP times to generate synthetic temporal scores and rescale them using the spatial loadings $Y_n$ to the $C$-channel space to form $\tilde{x}_n \in \mathbb{R}^{T' \times C}$. Use $G_{\hat{\phi}}$ fitted in Stage 3 to `EEGify' $\tilde{x}_n$ and thus obtain a final synthetic sample.

\section{Results} \label{sec:results}
\subsection{Datasets}

We examine the Children's Hospital Boston-Massachusetts Institute of Technology \citep[CHB-MIT;][]{shoeb2009application} and Siena scalp EEG datasets \citep{detti2020siena}. 
The CHB-MIT dataset comprises recordings from 22 pediatric subjects with intractable epilepsy, and the Siena dataset consists of recordings from 14 epileptic adults. In both sets of recordings, electrodes were positioned on the scalp according to the international 10-20 system. The CHB-MIT recordings have a consistent set of 18 bipolar channels in a double banana montage configuration, whereas the Siena dataset contains recordings using a monopolar montage, which we transformed into a double banana montage configuration. All recordings have a sampling frequency of 256 Hz (the Siena dataset was downsampled from 512 Hz). The datasets contain 198 and 47 expert-annotated seizure events, respectively, with CHB-MIT recordings ranging from 7 to 752 seconds and Siena recordings from 14 to 151 seconds (corresponding to $1.8\times 10^3- 1.9\times 10^6$ and $3.5\times 10^3- 3.8\times 10^4$ samples per channel). Due to our computational constraints, we remove patients 12 and 15 from the CHB-MIT dataset. 

We extract all seizure segments from the CHB-MIT and Siena datasets. The preprocessing was minimal, with only a 60 Hz and 50 Hz notch filter applied to the CHB-MIT and Siena datasets respectively to remove power line interference. The data is segmented into 1024 sample-long non-overlapping samples (4s at 256 Hz) and randomly select background segments to match the count of seizure segments for each patient. Those segments are normalized (across channels) and used to train the Conv-LSTM VAE and baseline models in a patient-dependent manner. Following training, we sample synthetic seizure segments from the generative models such that the number of original and synthetic seizure segments are identical for each patient.

\subsection{Baselines}

We benchmarked our method against three deep generative models: TimeVAE \cite{desai2021timevae}, COSCI-GAN \cite{seyfi2022generating}, and ImagenTime \cite{naiman2024utilizing}. TimeVAE and COSCI-GAN were selected based on recommendations from TSGBench \cite{tsgbench}, a comprehensive benchmark study that provides statistical analysis of generative model performance across diverse time series generation tasks. ImagenTime was selected as our third baseline due to its novel approach of transforming time series into images using invertible transforms, enabling the use of powerful diffusion-based vision models for generation.

TimeVAE was found to be particularly effective when using small-sized datasets, a critical consideration for patient-specific EEG generation. COSCI-GAN excelled at capturing complex multivariate relationships and explicitly models inter-channel dependencies through its cross-channel discriminator architecture, making it well-suited for preserving the spatial correlations inherent in multi-channel EEG recordings. ImagenTime achieved state-of-the-art results across long, and ultra-long sequences in unconditional generation tasks, demonstrating particular strength in capturing complex temporal dynamics through its diffusion backbone. All the models were trained in a patient-dependent manner to match with our method. Additional training parameters for all the baselines are detailed in the Appendix \ref{app_sec:baseline_hyperparams}.

\subsection{Feature-based evaluation}

\begin{figure*}[t!]
    \centering
    \includegraphics[width=0.95\textwidth]{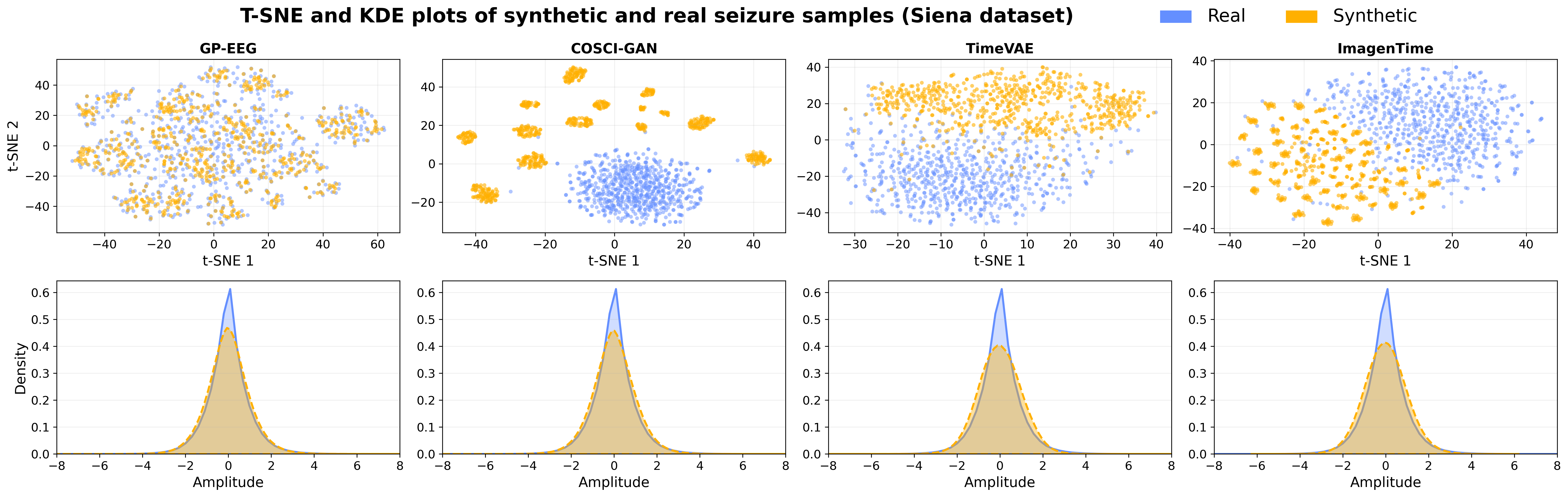}\\
    \vspace{0.5cm}
    \includegraphics[width=0.95\textwidth]{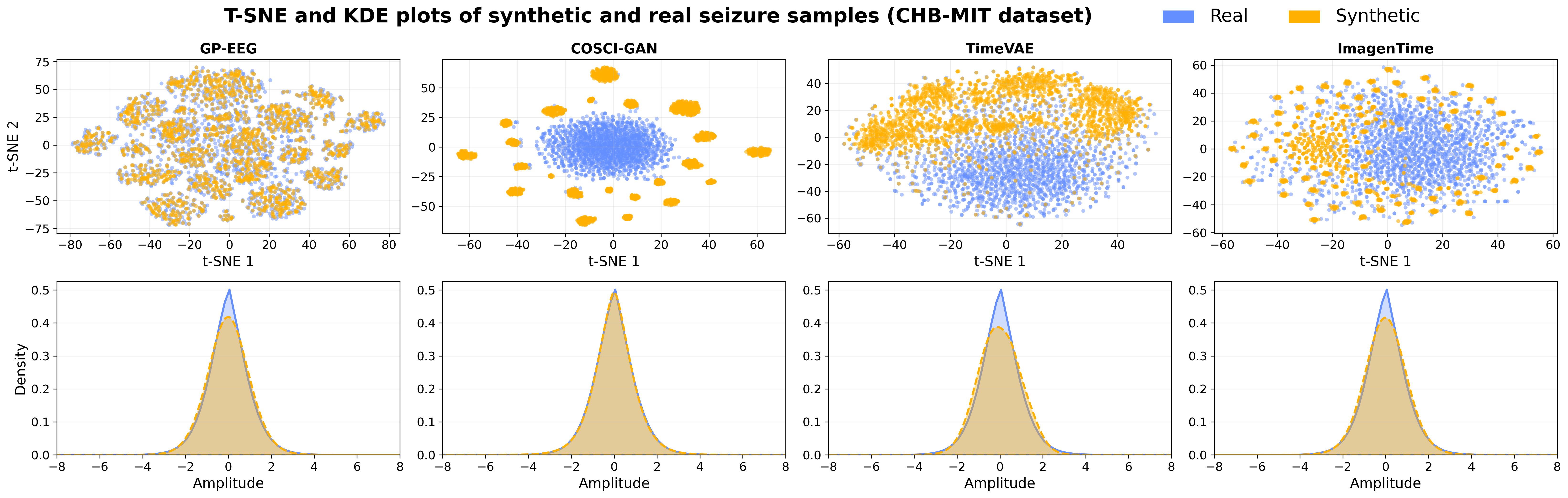}
    \caption{Qualitative comparison of real and synthetic seizure data on the Siena (top) and CHB-MIT (bottom) datasets. For each dataset, the upper row shows t-SNE embeddings (perplexity=30) and the lower row shows KDE of amplitude distributions.}
    \label{fig:tsne_kde}
\end{figure*}

We quantitatively evaluate the quality of synthetic EEG samples using four metrics commonly used in the literature: the Marginal Distribution Difference (MDD), the Autocorrelation Difference (ACD), the Skewness Difference (SD) and the Kurtosis Difference (KD) \cite{tsgbench}. 

\textbf{MDD} measures how closely the amplitude distributions of generated data match those of real data. For each channel and time step, we compute empirical histograms over all samples and report the mean absolute difference across bins between the real and generated distributions, averaging across all channels, time steps, and bins. See Appendix \ref{app_sec:mdd_metric} for mathematical details.

\textbf{ACD} measures the discrepancy in temporal dependency structure between real and generated sequences. We compute the autocorrelation function up to lag $L=63$ for each channel, then report the $2-$norm norm of the difference: 
\[
\text{ACD} = \sqrt{\sum_{c=1}^{C} \sum_{l=1}^{L} \left( \rho^{(c)}(l) - \tilde{\rho}^{(c)}(l) \right)^2},\] 
where $\rho^{(c)}(l)$ and $\tilde{\rho}^{(c)}(l)$ are the autocorrelation at lag $l$ for channel $c$ of the real and generated data, respectively. Lower ACD indicates that the generated data better preserves the temporal dynamics of the original signal, such as rhythmic structure and short-term dependencies.

\textbf{SD and KD} measure discrepancies in the third and fourth standardised moments respectively, capturing differences in distribution asymmetry and tail behavior between real and generated data. For real and generated time series $x$ and $\tilde{x}$, define their means and standard deviations as $\mu, \sigma$ and $\tilde{\mu}, \tilde{\sigma}$. We define
\[
\left|\frac{\mathbb{E}\{(x - \mu)^p\}}{\sigma^p} - \frac{\mathbb{E}\{(\tilde{x} - \tilde{\mu})^p\}}{\tilde{\sigma}^p}\right|,
\]
which takes the value of SD and KD for $p=3$ and $p=4$, respectively. The results are shown in Table \ref{tab:metrics_results}. Additional box plots depicting the distribution of Catch22 and EEG-based features between the real and synthetic data are provided in Appendix \ref{app_sec:feature_space}.

\begin{table}[h]
\caption{Quantitative comparison of generative methods on the Siena and CHB-MIT datasets. Lower values indicate better performance. Best results in \textbf{bold}.}
\label{tab:metrics_results}
\begin{center}
\begin{small}
\begin{tabular}{lcccc}
\toprule
\multicolumn{5}{c}{\textbf{Siena}} \\
\midrule
Method & MDD $\downarrow$ & ACD $\downarrow$ & SD $\downarrow$ & KD $\downarrow$ \\
\midrule
COSCI-GAN     & 0.289 & 3.300 & 0.510 & 6.342 \\
TimeVAE       & 0.261 & 4.916 & 0.511 & 6.064 \\
ImagenTime    & 0.263 & 1.871 & 0.411 & 6.598 \\
GP-EEG (ours) & \textbf{0.251} & \textbf{0.341} & \textbf{0.326} & \textbf{5.316} \\
\midrule
\multicolumn{5}{c}{\textbf{CHB-MIT}} \\
\midrule
Method & MDD $\downarrow$ & ACD $\downarrow$ & SD $\downarrow$ & KD $\downarrow$ \\
\midrule
COSCI-GAN     & 0.245 & 1.265 & 0.312 & \textbf{1.562} \\
TimeVAE       & 0.195 & 5.299 & 0.301 & 2.006 \\
ImagenTime    & 0.197 & 1.323 & 0.255 & 1.822 \\
GP-EEG (ours)  & \textbf{0.183} & \textbf{0.304} & \textbf{0.163} & 1.696 \\
\bottomrule
\end{tabular}
\end{small}
\end{center}
\end{table}

\subsection{Visual and qualitative evaluation}

The t-SNE and kernel density estimates (KDE) plots of synthetic and real EEG seizure segments for the Siena and CHB-MIT datasets are shown in Figure \ref{fig:tsne_kde}. GP-EEG exhibits the closest overlap with real data in both embedding space and amplitude distribution, with the exception of the CHB-MIT amplitude KDE, where COSCI-GAN shows a slightly better performance. Time-domain plots of synthetic samples are presented in Appendix \ref{app:example_seizures}.

\subsection{Train-on-synthetic, test-on-real}

For the model-based evaluation of the synthetic data, we propose a variant of the standard train-on-synthetic, test-on-real (TSTR) scheme. Specifically, we train the model using both synthetic seizures and real background EEG samples. The trained model is then evaluated on real background and seizure EEG samples. This setup represents a more challenging scenario compared to training solely on synthetic data, as the network may exploit domain shift during training rather than learn meaningful differences between seizure and background samples. We adopt EEGNet4,2 for all experiments \cite{lawhern2018eegnet}, with training details provided in Appendix \ref{app:tstr}. Evaluation is performed using a leave-one-patient-out (LOPO) cross-validation strategy, where the test set consists exclusively of real EEG seizure and background segments from the held-out patient. Results are reported as the average difference in performance across folds between the baseline and TSTR scheme. Specifically, we subtract each performance metric for the baseline from that of the TSTR scheme. Results are shown in Table \ref{tab:metrics_tstr}.  The full model performance is reported in Appendix in Table \ref{app_tab:metrics_tstr}. Note that all results for the baselines are large negative values, reflecting that the models learned to discriminate between real and synthetic samples, rather than training for seizure detection. Comparatively, GP-EEG has vastly superior performance.

\begin{table}[h]
\caption{Average change in performance between baseline and TSTR scheme (TSTR minus baseline). Best results in \textbf{bold}.}
\label{tab:metrics_tstr}
\begin{center}
\begin{small}
\begin{tabular}{lcccc}
\toprule
\multicolumn{5}{c}{\textbf{Siena}} \\
\midrule
Method & Acc (\%) & Pre (\%) & Rec (\%) & F1 (\%)\\
\midrule
COSCI-GAN     & -25.07 & -66.48 & -70.98 & -71.00 \\
TimeVAE       & -24.51 & -77.20 &  -71.31 & -71.64\\
ImagenTime    & -23.77 &-41.48 & -69.61  & -68.42 \\
GP-EEG (ours) & \textbf{0.32} & \textbf{6.46} & \textbf{-8.45} & \textbf{-3.05} \\
\midrule
\multicolumn{5}{c}{\textbf{CHB-MIT}} \\
\midrule
Method & Acc (\%) & Pre (\%) & Rec (\%) & F1 (\%)\\
\midrule
COSCI-GAN     & -21.70 & -74.23 & -64.57 & -67.54 \\
TimeVAE       & -21.65 & -69.68 & -64.50 & -67.41 \\
ImagenTime    & -21.67 & -69.68 & -64.50 & -67.41 \\
GP-EEG (ours) & \textbf{-5.75} & \textbf{-3.2} & \textbf{-13.35} & \textbf{-9.79} \\
\bottomrule
\end{tabular}
\end{small}
\end{center}
\end{table}

\subsection{Data augmentation evaluation}\label{sec:data_augmentation_eval}

We evaluate the effectiveness of the proposed synthetic seizure data as a data augmentation tool for patient-independent seizure detection. Unlike the TSTR setting, which assesses how realistic the synthetic data is in isolation, this evaluation measures whether augmenting the dataset with synthetic samples can improve model performance. Evaluation is performed using a LOPO cross-validation strategy, where the test set consists exclusively of real EEG seizure and background segments from the held-out patient.

All experiments employ the EEGNet4,2 architecture \cite{lawhern2018eegnet}. Additional details about the training can be found in Appendix \ref{app:augmentation_eval}. Results are reported by providing the augmented performance minus the baseline performance for each metric averaged across folds, and can be seen in Table \ref{tab:metrics_augmentation}. The full model performance is reported in Appendix in Table \ref{app_tab:metrics_augmentation}. 

\begin{table}[h]
\caption{Average change in performance across folds between the baseline and augmented experiment (augmented minus baseline). Best results in \textbf{bold}.}
\label{tab:metrics_augmentation}
\begin{center}
\begin{small}
\begin{tabular}{lcccc}
\toprule
\multicolumn{5}{c}{\textbf{Siena}} \\
\midrule
Method & Acc (\%) & Pre (\%) & Rec (\%) & F1 (\%)\\
\midrule
COSCI-GAN     & 2.95 & \textbf{10.75} & -7.01 & -0.54\\
TimeVAE       & -3.08 & 0.12 & -19.44 & -12.70\\
ImagenTime    & 1.70 & 6.29 & -5.47 & -0.27 \\
GP-EEG (ours) & \textbf{4.93} & 8.70 & \textbf{1.25} & \textbf{5.11} \\
\midrule
\multicolumn{5}{c}{\textbf{CHB-MIT}} \\
\midrule
Method & Acc (\%) & Pre (\%) & Rec (\%) & F1 (\%)\\
\midrule
COSCI-GAN     & -3.01 & 6.58 & -17.23 & -10.52\\
TimeVAE       & -3.99 & \textbf{10.17} & -22.31 & -14.23\\
ImagenTime    & -0.89 & 3.68 & -8.16 & -3.70\\
GP-EEG (ours) & \textbf{1.75} & 2.37 & \textbf{2.26} & \textbf{2.75}\\
\bottomrule
\end{tabular}
\end{small}
\end{center}
\end{table}

\section{Conclusion} \label{sec:conclusion}

We have presented a novel framework for generating synthetic ultra-long, multivariate epileptic EEG recordings that addresses fundamental challenges in clinical neurophysiology data synthesis. Our approach generates patient-specific 18-channel epileptic EEG recordings while maintaining interpretability at each stage of the pipeline, and achieves state-of-the-art performance on both evaluated datasets. To the best of our knowledge, this is the first study to produce interpretable, multivariate, non-stationary, long-horizon EEG seizure data.

The primary computational bottleneck of our method lies in the GP regression and KL divergence calculations for kernel parameter discretization - both of which can be parallelized. In addition, our approach is fully interpretable, enables synthesis of realistic ultra-long recordings from limited training data, and leverages SVD decomposition as a computationally efficient strategy for preserving meaningful inter-channel correlations, while reducing dimensionality for subsequent generative modeling stages.

Several future extensions for this work are possible. First, the current patient-specific discretization of kernel hyperparameters could be generalized to a cross-patient framework by computing the KL divergence between the sets of patient-specific discretized parameters (a more computationally tractable methodology than computing the KL divergence between all kernel parameters across all patients). Second, the GP kernel selection and hyperparameter grid search could be enhanced through automated approaches such as those proposed in recent work on time series forecasting \cite{abdessalem2017automatic}, which would become particularly advantageous given access to larger training datasets. 

Our work demonstrates that classical mathematical tools, when thoughtfully combined with modern machine learning components, can tackle challenging generative tasks where end-to-end deep learning approaches face constraints, especially when dealing with high dimensional, but relatively small, non-stationary, datasets.

\section*{Impact Statement}

This work develops an interpretable pipeline for generating long-horizon, multi-channel synthetic EEG recordings. The primary positive impact is to support methodological research in a clinically important but data-scarce settings by enabling augmentation, stress-testing, and benchmarking of learning algorithms when real recordings are limited, expensive to curate, or difficult to share. In particular, improved training and evaluation of seizure-detection models has the potential to yield downstream medical benefits, including more reliable seizure monitoring and their earlier detection, which could support clinical decision-making and patient safety in real-world healthcare. Synthetic generation may also facilitate reproducibility and collaboration by reducing reliance on direct redistribution of sensitive patient data.

Potential negative impacts include misuse or over-interpretation of synthetic EEG as clinically valid evidence, deployment of models trained on synthetic data without adequate validation on real patient cohorts, and amplification of biases if the training datasets are unrepresentative of the broader patient population. In addition, although synthetic data can reduce disclosure risk, generative models may still leak information about individuals if trained or released without appropriate safeguards.

To mitigate these risks, our intended use is research and development rather than clinical decision-making. We recommend clearly labeling synthetic data, reporting fidelity metrics alongside limitations, evaluating privacy leakage risks where appropriate, and validating downstream models on held-out real data across diverse cohorts before any clinical use. We also encourage adherence to relevant data governance and ethical review procedures when training on clinical EEG datasets.

\bibliography{references}
\bibliographystyle{icml2026}

\newpage
\appendix
\onecolumn
\section{Example synthetic seizures produced by GP-EEG, COSCI-GAN, TimeVAE and, ImagenTime} \label{app:example_seizures}

\begin{figure}[h!]
    \centering
    \includegraphics[width=0.90\textwidth]{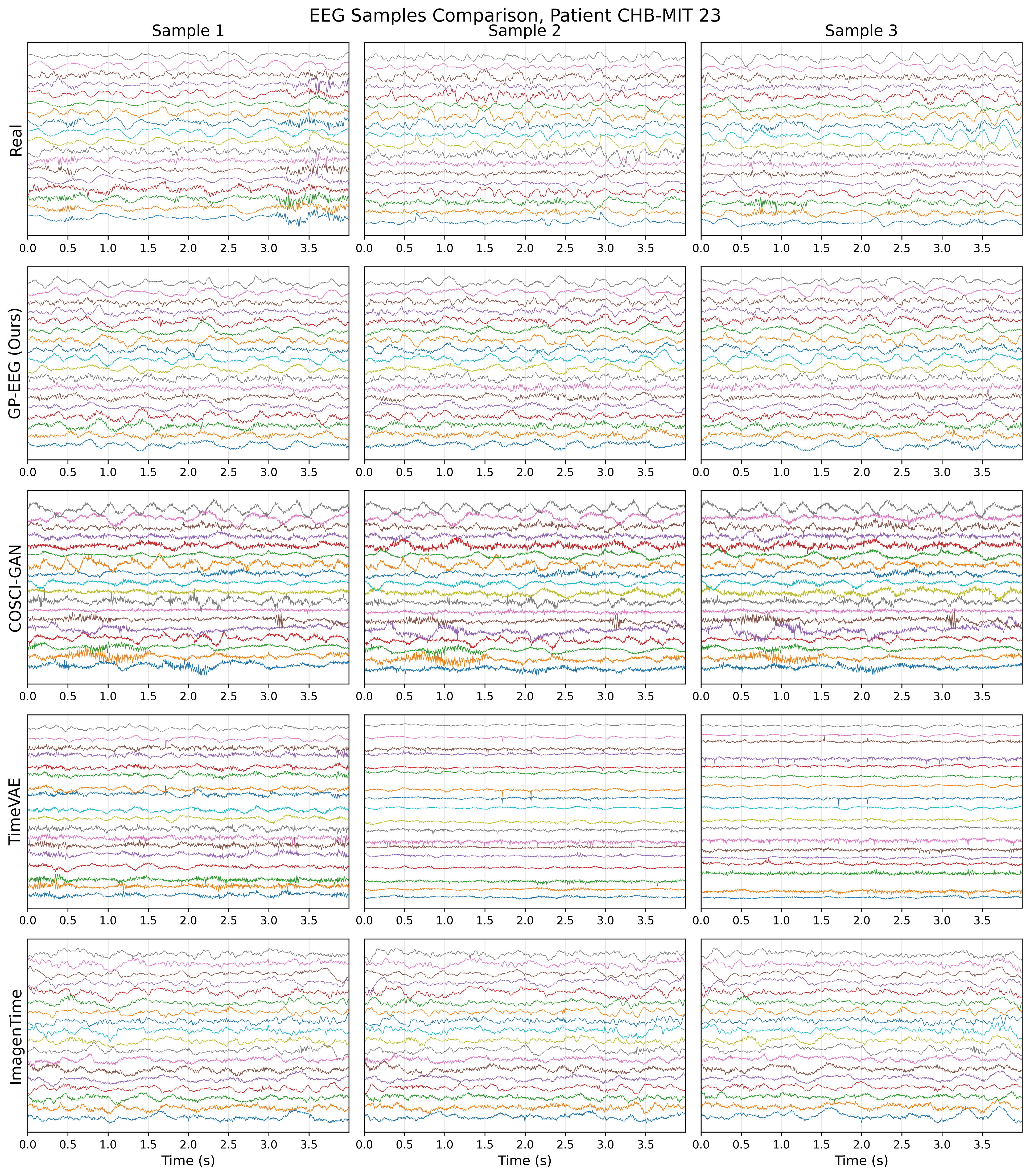}
    \caption{Sample seizures on a 4 second interval. The first row is a real seizure segment from patient 23 of the CHB-MIT dataset, and rows 2, 3 and 4 are synthetic output produced by GP-EEG, COSCI-GAN, TimeVAE, and ImagenTime. Each column is a different realization.}
    \label{fig:comparison_algo_ts}
\end{figure}

\begin{figure}[h!]
    \centering
    \includegraphics[width=0.90\textwidth]{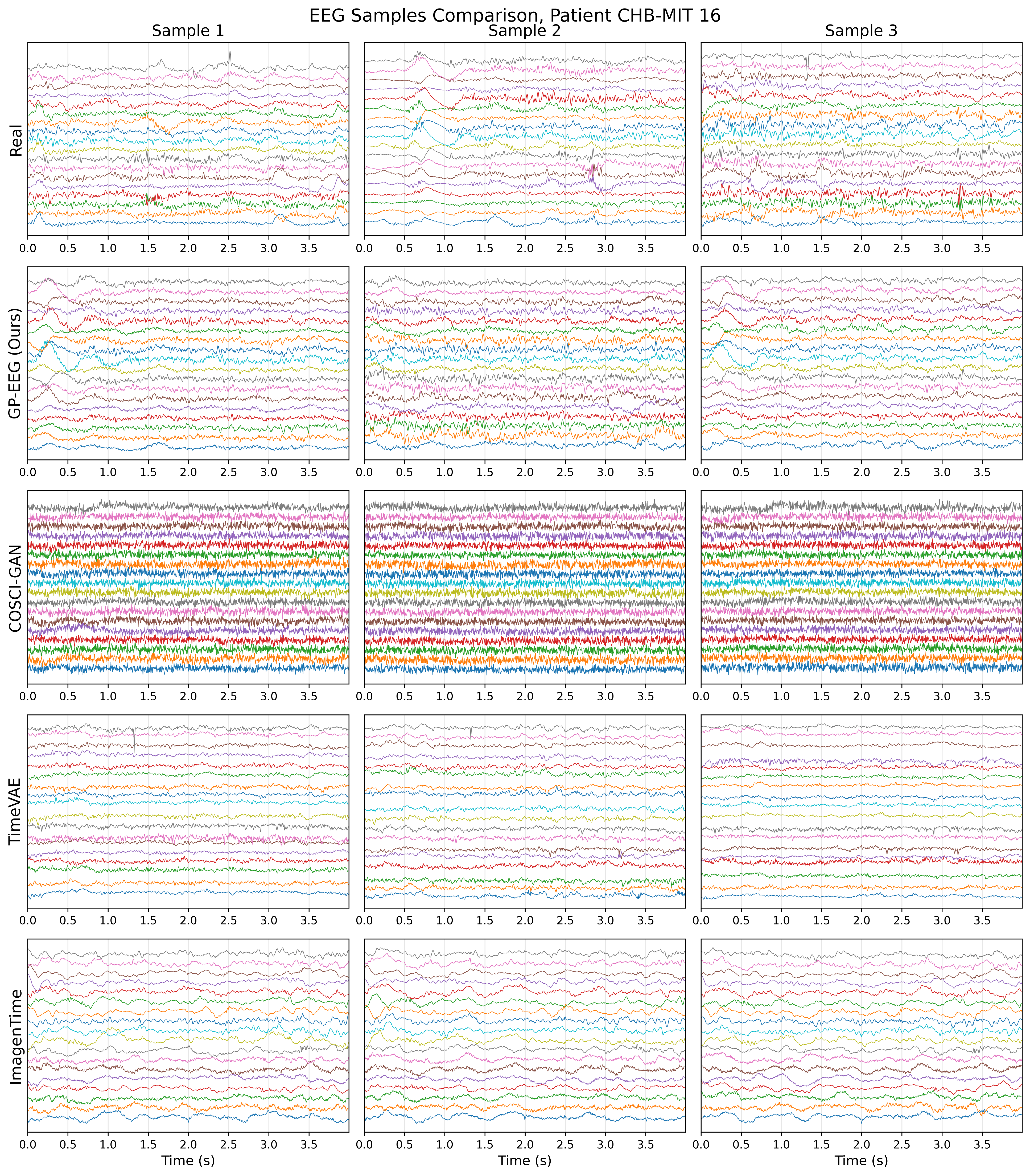}
    \caption{Sample seizures on a 4 second interval. The first row is a real seizure segment from patient 16 of the CHB-MIT dataset, and rows 2, 3 and 4 are synthetic output produced by GP-EEG, COSCI-GAN, TimeVAE, and ImagenTime. Each column is a different realization.}
    \label{fig:comparison_algo_ts_2}
\end{figure}

\begin{figure}[h]
    \centering
    \includegraphics[width=\textwidth]{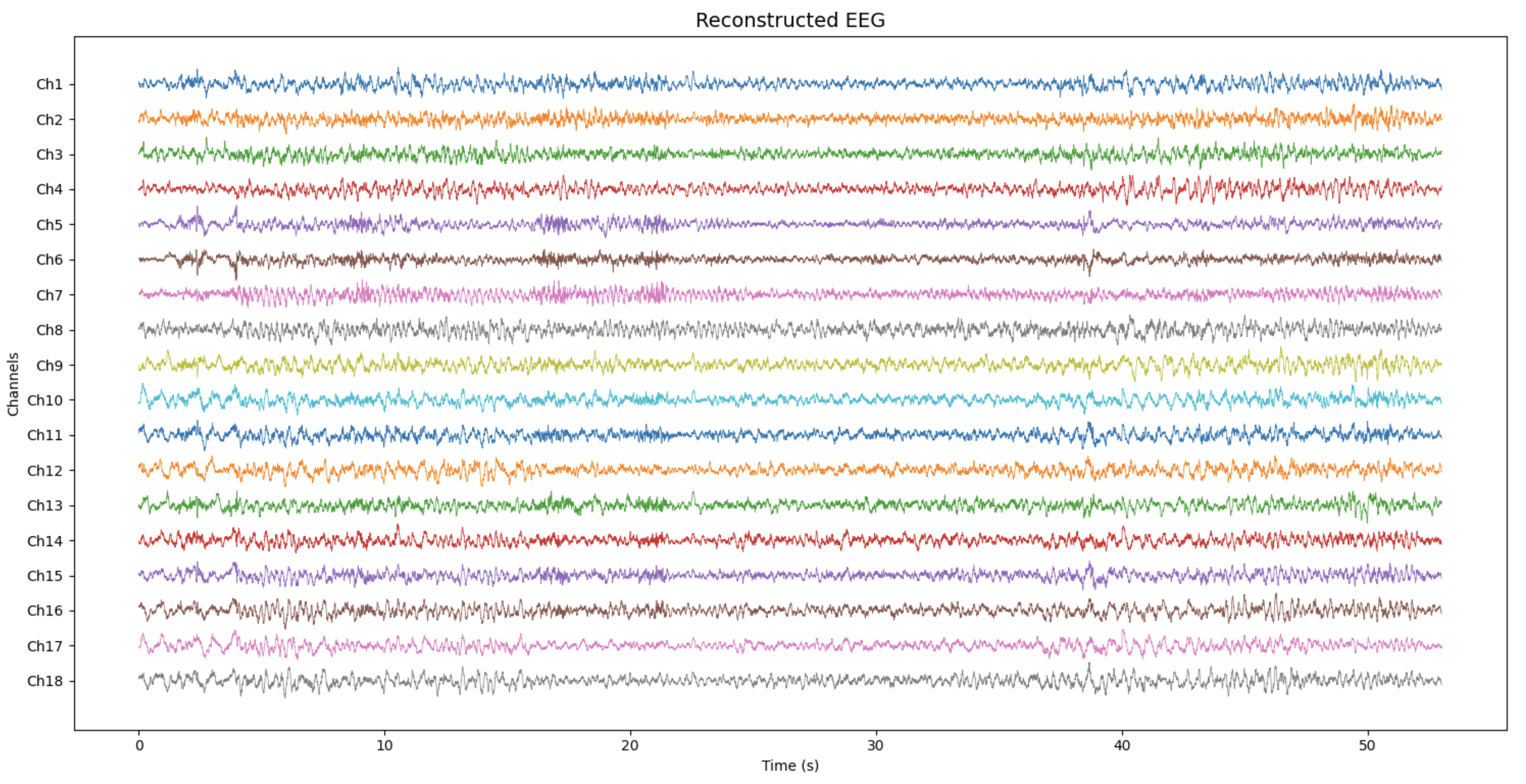}
    \caption{Sample synthetic seizure segment, obtained using GP-EEG, on a 53 seconds interval.}
    \label{fig:full_sz}
\end{figure}

In this section, we provide example output from the different models discussed in the main body when trained on patient 23 and 16 of the CHB-MIT dataset. We provide three real seizure samples and three synthetic seizures produced by GP-EEG, COSCI-GAN, TimeVAE, and ImagenTime. Figures \ref{fig:comparison_algo_ts} and \ref{fig:comparison_algo_ts_2} correspond to patients 23 and 16 of CHB-MIT, respectively. For both patients, GP-EEG produces synthetic seizures that closely resemble the original recordings, preserving realistic inter-channel correlations, amplitude fluctuations, and temporal dynamics across samples. Figure \ref{fig:full_sz} depicts a non-truncated synthetic seizure sample using GP-EEG on a length of 53 seconds.


\clearpage
\section{Run-time analysis} \label{app:run_time_analysis}

In this section, we provide output of the run-time of our algorithm using the Siena dataset as seen in Table \ref{tab:patient_files_siena}. 

\begin{table}[!h]
\centering
\begin{threeparttable}
\caption{Patient-wise file ID mapping for the Siena dataset, and associated run-time for GP regression and KL computations.}
\label{tab:patient_files_siena}
\begin{tabular}{lcccccc}
\toprule
\textbf{Patient ID} & \textbf{File IDs} & \makecell{\textbf{Number of}\\\textbf{segments}} & \makecell{\textbf{GP regression}\\\textbf{time (hh:mm)}} & \makecell{\textbf{Number of KL}\\\textbf{computations}} & \makecell{\textbf{KL run time (hh:mm:ss)}\\\textbf{(CPU time use)}} & \makecell{\textbf{KL walltime}\\\textbf{(hh:mm:ss)}} \\
\midrule
PN 00 & 42--46  & 591 & 37:30 & 348690 & 251:07:57 & 03:42:19\\ 
PN 01 & 0--1  & 115 & 16:18 & 74802 & 43:57:57 & 03:01:46\\
PN 03 & 25--26  & 370 & 24:58  & 136530 & 78:34:59 & 03:00:12 \\
PN 05 & 27--29  & 162 & 11:00 & 26082 & 15:12:01 & 02:57:43 \\
PN 06 & 30--34  & 448 & 30:24 & 200256 & 125:22:18 & 03:11:22\\
PN 07 & 19--19  & 93 & 06:28 & 8556 & 04:49:22 & 02:51:20 \\
PN 09 & 20--22  & 299 & 20:43 & 89102 & 50:54:00 & 02:55:43 \\
PN 10 & 4--13 & 660 & 41:16 & 434940 & 340:46:24 & 04:00:57 \\
PN 11 & 14--14  & 99 & 06:20 & 9702 & 05:32:15 & 02:52:11 \\
PN 12 & 15--18  & 555 & 34:37 & 307470 &  214:30:12 &  03:35:09 \\
PN 13 & 35--37  & 469 & 31:15 & 219492 & 141:04:37 & 03:18:21 \\
PN 14 & 38--41  & 199 & 14:59 & 39402 & 22:47:00 & 02:56:27\\
PN 16 & 2--3  & 322 & 23:06 & 103362 & 61:08:17 & 03:01:05 \\
PN 17 & 23--24  & 272 & 16:50 & 73712 & 41:33:32 & 02:52:50\\
\midrule
TOTAL & 0-46 & 4654 & $\approx$ 315h & 2072098 & 1397:20:51 & 44:17:25 \\
\bottomrule
\end{tabular}
\end{threeparttable}
\end{table}

\section{Dimensionality reduction with SVD/PCA}\label{app:dimred_svd_pca}
In this section, we provide a brief overview of dimensionality reduction using SVD/PCA. This is an important part of the GP-EEG algorithm.

Multichannel EEG recordings often contain substantial redundancy across channels: many electrodes reflect mixtures of a smaller number of latent sources (oscillatory activity, transients, artefacts), observed through subject-specific spatial mixing. This motivates a low-rank approximation in which a $C$-channel recording can be represented by only $d\ll C$ latent time series together with $d$ spatial patterns. In our setting, this compression is performed with singular value decomposition (SVD), which is closely related to principal component analysis (PCA).

\paragraph{SVD as an optimal low-rank approximation.}
For a column-centered matrix $x_n\in\mathbb{R}^{T_n\times C}$ (constructed as in the main text), the SVD expresses $x_n$ as
\begin{equation}\label{eq:app_svd_full}
x_n \;=\; \sum_{r=1}^{\mathrm{rank}(x_n)} \sigma_{n,r}\, u_{n,r}\, v_{n,r}^\top,
\end{equation}
where $\sigma_{n,1}\ge \sigma_{n,2}\ge \cdots \ge 0$ are the singular values, $\{u_{n,r}\}$ are orthonormal vectors in $\mathbb{R}^{T_n}$, and $\{v_{n,r}\}$ are orthonormal vectors in $\mathbb{R}^{C}$. Truncating this expansion to its first $d$ terms yields the rank-$d$ approximation
\begin{equation}\label{eq:app_svd_trunc}
x_{n,d} \;:=\; \sum_{r=1}^{d} \sigma_{n,r}\, u_{n,r}\, v_{n,r}^\top,
\end{equation}
which provides the best rank-$d$ approximation to $x_n$ in squared Frobenius norm:
\begin{equation}\label{eq:app_ey}
x_{n,d}\in \mathrm{argmin}_{\mathrm{rank}(Z)\le d}\;\|x_n - Z\|_F^2.
\end{equation}
Thus, retaining only the leading $d$ components produces an optimally compressed representation of the multichannel signal among all linear rank-$d$ representations.

\paragraph{Connection to PCA across channels.}
Because $x_n$ is column-centered, the right singular vectors correspond to principal directions of the empirical channel covariance. Specifically,
\begin{equation}\label{eq:app_pca_cov}
\frac{1}{T_n}\,x_n^\top x_n
\;=\;
V_n \left(\frac{D_n^2}{T_n}\right) V_n^\top,
\end{equation}
so the columns of $V_n$ are the PCA loading vectors and the eigenvalues are $\{\sigma_{n,r}^2/T_n\}$. 

\paragraph{Relevance for long-horizon generation.}
The low-rank structure implied by \eqref{eq:app_svd_trunc} reduces the generative task from modeling $C$ coupled channels to modeling only $d$ latent temporal degrees of freedom, which is substantially more tractable for long and ultra-long sequences. Moreover, reconstruction through the learned spatial loadings preserves realistic multivariate structure: cross-channel dependencies are enforced by the shared low-rank representation rather than being learned implicitly by a high-capacity end-to-end generator. In our pipeline, this separation is particularly useful because subsequent stages operate on the low-dimensional temporal scores while retaining patient-specific spatial structure through the corresponding spatial loadings.

\begin{figure}[h!]
    \centering
    \includegraphics[width=\textwidth]{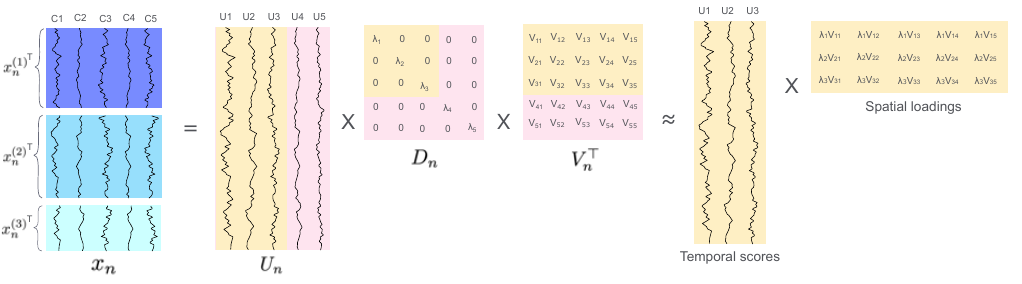}
    \caption{SVD of stacked EEG recordings, for patient $n$. The matrix $x_n$ contains multiple vertically stacked EEG segments, the mean of each column is set to zero. $x_n$  is decomposed into temporal scores $U_n$, singular values $D_n$, and spatial loadings $V_n^T$. The reconstruction using a subset of singular vectors demonstrates how the original 18-channel EEG data can be represented as a linear combination of orthogonal temporal scores weighted by spatial loadings.}
    \label{fig:appendix_svd}
\end{figure}

In Figure \ref{fig:appendix_svd}, we provide a cartoon representation of how the samples for a patient are stacked and decomposed. We see the approximate recovery by selecting $d=3$ and splitting the data into the product of temporal scores and spatial loadings.

\section{Gaussian process regression} \label{app:GP_regression}

\begin{figure}[h!]
    \centering
    \begin{subfigure}{0.47\textwidth}
        \centering
        \includegraphics[width=\textwidth]{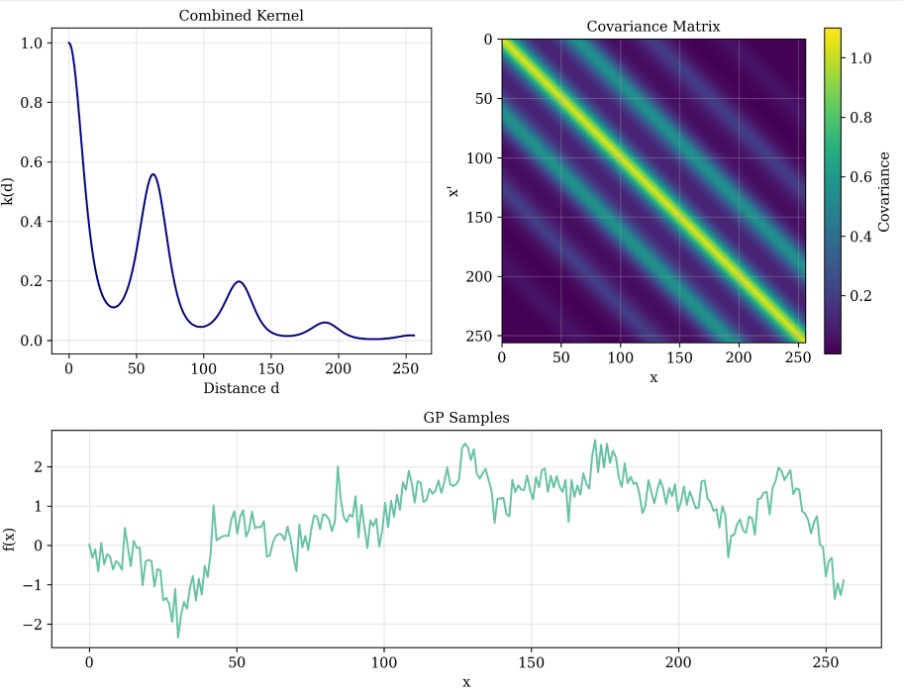}
        \caption{$\ell_m=73$}
        \label{fig:gp_params_a}
    \end{subfigure}%
    \hfill
    \begin{subfigure}{0.47\textwidth}
        \centering
        \includegraphics[width=\textwidth]{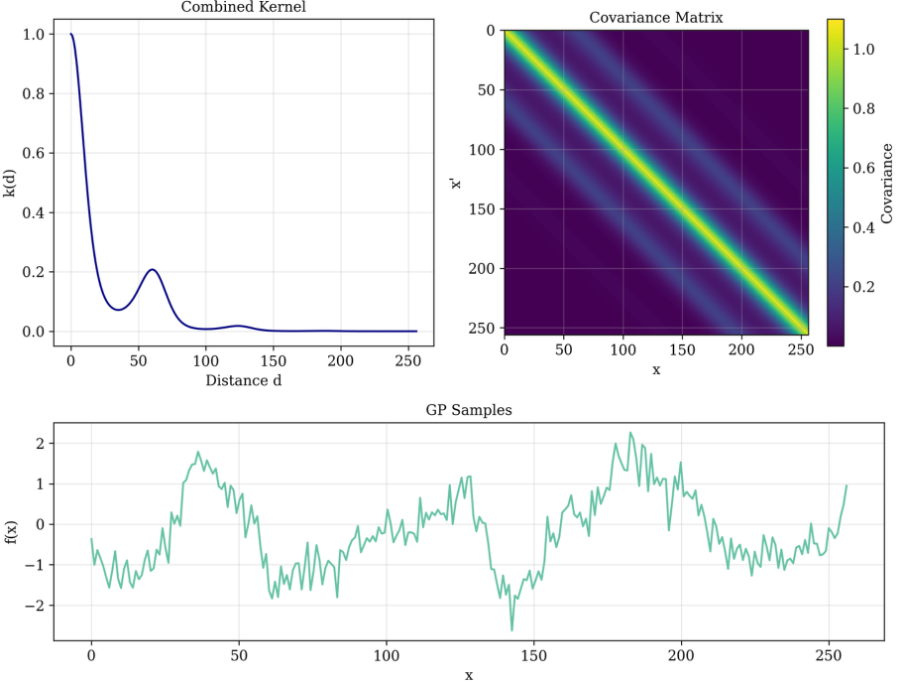}
        \caption{$\ell_m=36.5$}
        \label{fig:gp_params_b}
    \end{subfigure}
    
    \vspace{0.25cm}
    
    \begin{subfigure}{0.47\textwidth}
        \centering
        \includegraphics[width=\textwidth]{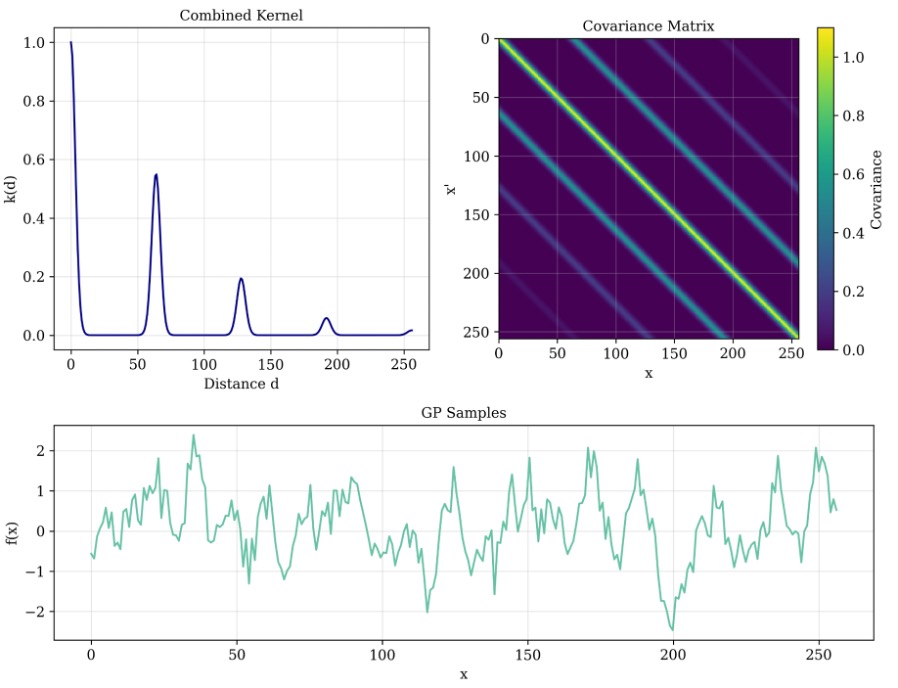}
        \caption{$\ell_p=0.1$}
        \label{fig:gp_params_c}
    \end{subfigure}%
    \hfill
    \begin{subfigure}{0.47\textwidth}
        \centering
        \includegraphics[width=\textwidth]{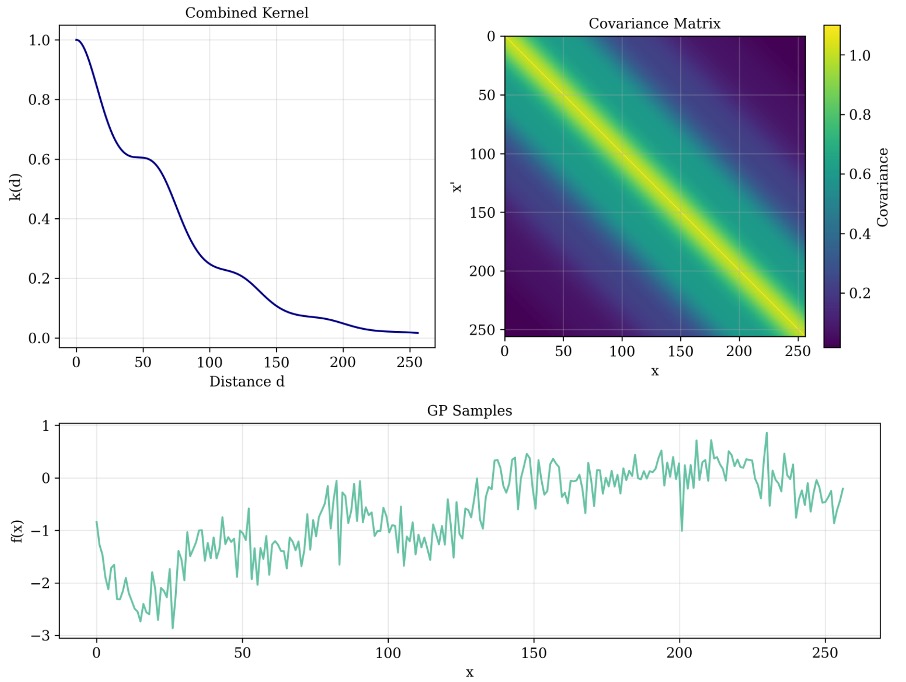}
        \caption{$\ell_p=8$}
        \label{fig:gp_params_d}
    \end{subfigure}
    
    \vspace{0.25cm}
    
    \begin{subfigure}{0.47\textwidth}
        \centering
        \includegraphics[width=\textwidth]{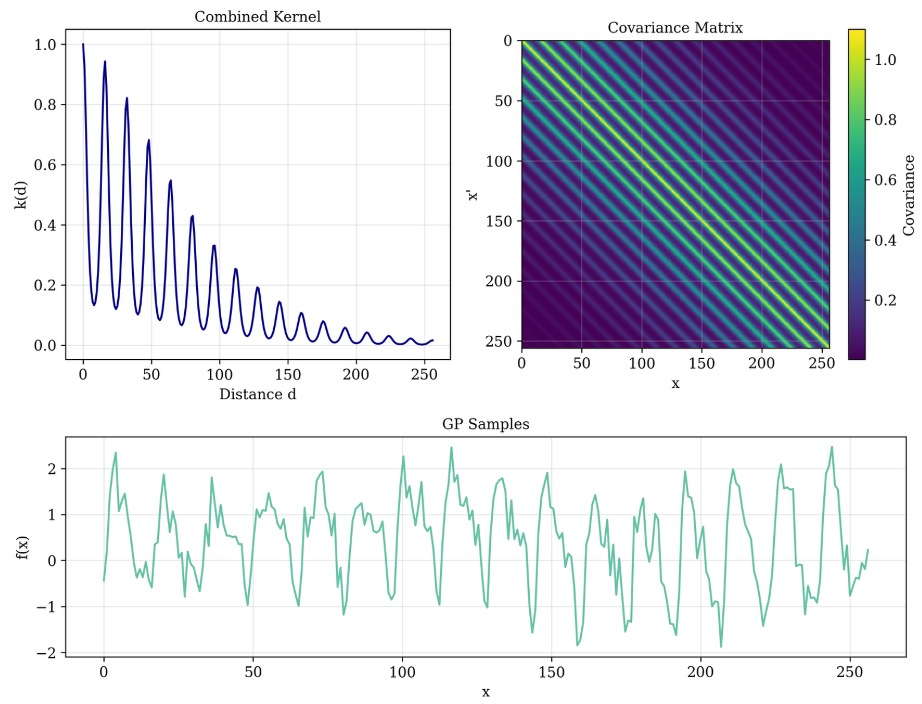}
        \caption{$p=16$ samples}
        \label{fig:gp_params_e}
    \end{subfigure}%
    \hfill
    \begin{subfigure}{0.47\textwidth}
        \centering
        \includegraphics[width=\textwidth]{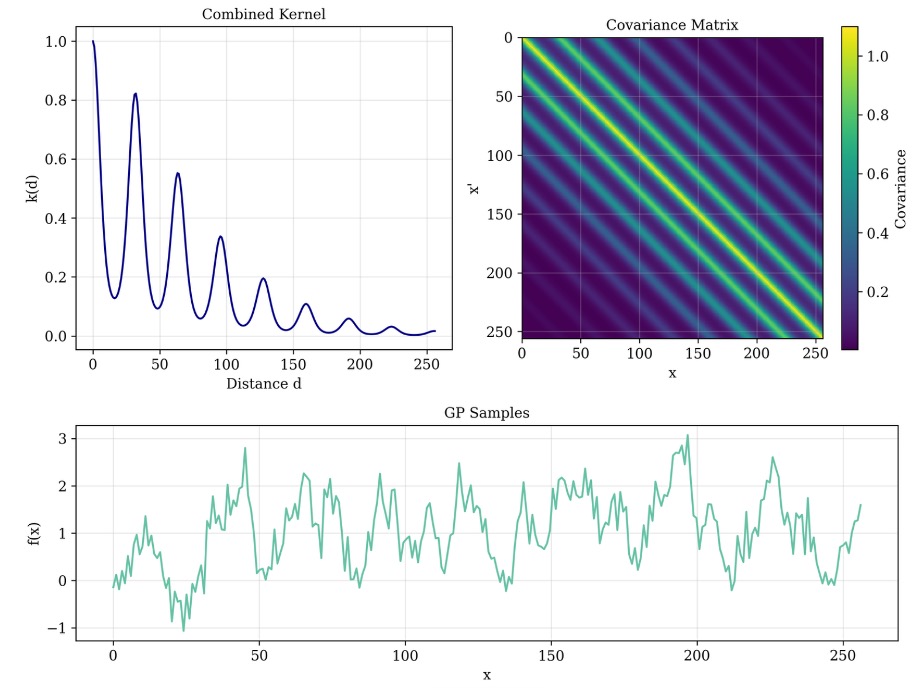}
        \caption{$p=32$ samples}
        \label{fig:gp_params_f}
    \end{subfigure}
    
    \caption{Effect of kernel hyperparameters on GP samples using the kernel described in Equation \ref{eq:kernel}. \textbf{(a-b)} Varying Matérn lengthscale $\ell_m$ controls the overall smoothness, with larger values producing smoother functions ($\ell_p=1$, $p=256/4$). \textbf{(c-d)} Varying periodic lengthscale $\ell_p$ controls the regularity of oscillations, with larger values allowing greater deviation from strict periodicity ($\ell_m=73$, $p=256/4$). \textbf{(e-f)} Varying period $p$ controls the frequency of oscillations ($\ell_m=73$, $\ell_p=1$). Other parameters are held constant at $\sigma_f^2=1$ and $\sigma_n^2=0.1$.}
    \label{fig:gp_kernel_params}
\end{figure}

This section provides a description of the kernel used for the GP regression, details on how the hyperparameters for the kernel were selected and specifics on the \textit{Python} implementation.

\subsection{Choice of kernel} \label{app:choice_of_kernel}
A \emph{Gaussian process} (GP) is a distribution over functions: it is a collection of random variables $\{f(x) : x\in\mathcal{X}\}$ such that any finite sub-collection has a joint multivariate Gaussian distribution. We write
\begin{equation}
    f(x) \sim \mathcal{GP}\!\big(m(x),\, k(x,x')\big),
\end{equation}
where the mean function $m(x) = \mathbb{E}\!\left[f(x)\right]$ specifies the prior expected value at input $x$, and the covariance (kernel) function
\begin{equation}
    k(x,x')=\mathrm{Cov}\!\big(f(x),f(x')\big)
    =\mathbb{E}\!\left[(f(x)-m(x))(f(x')-m(x'))\right]
\end{equation}
encodes assumptions about smoothness, periodicity, and correlation structure of sample paths. It is common to set $m(x)\equiv 0$ without loss of generality when data are centered.

In this work we apply Gaussian process regression to model single-channel EEG segments as a function of time, so the kernel is a function of temporal separation $r=\|x-x'\|$ (with $x,x'\in\mathbb{R}$ denoting time). To capture approximately periodic oscillations whose similarity decays over time---a common pattern in quasi-stationary EEG sub-segments---we use a periodic kernel \cite{nicholson2022quasi}. Specifically, we multiply a periodic kernel by a Mat\'ern-$3/2$ kernel:
\begin{equation}
    k(x,x')
    = \sigma_f^2
    \exp\!\left(
        -\frac{2}{\ell_p^2}\sin^2\!\left(\pi\frac{\|x-x'\|}{p}\right)
    \right)
    \left(
        1+\frac{\sqrt{3}\|x-x'\|}{\ell_m}
    \right)
    \exp\!\left(
        -\frac{\sqrt{3}\|x-x'\|}{\ell_m}
    \right),
    \label{app_eqn:kernel}
\end{equation}
where $\sigma_f^2$ controls the marginal variance, $p$ is the (approximate) period, $\ell_p$ controls how strictly periodic the function is, and $\ell_m$ governs the timescale over which correlations decay. The Mat\'ern kernel reduces the smoothness of the samples, thus making them more EEG-like in appearance.

The kernel in \eqref{app_eqn:kernel} gives four hyperparameters to select: $\theta = (\sigma_f, \ell_p, \ell_m, p)$. The length scale of the Mat\'ern kernel $\ell_m$ controls the distance over which neighboring points have a strong effect on one another. A small $\ell_m$ corresponds to a small correlation between inputs, and vice versa for large values. Similarly, the length scale of the periodic kernel $\ell_p$ determines the shape of the periodic component and how quickly the correlation decays. The period $p$ defines the distance between repetitions of the function, representing the fundamental frequency of the data. The signal variance $\sigma_f^2$ provides information regarding the amplitude of the function being modeled. A large $\sigma_f^2$ will result in a function with a very large amplitude. 

In Figure \ref{fig:gp_kernel_params}, we show plots of the kernel for different hyperparameter values, as well as samples from a GP with that hyperparameter combination. We vary the length scales and the period, see observe how the behavior of the kernel differs in each case. In each subplot, there is a plot of the kernel value as a function of distance, a plot of the kernel on a 2D grid, and a sample realization.

\subsection{Hyperparameter grid search} \label{app:hyperparam_grid_search}

The marginal likelihood surface for a Gaussian process is highly non-convex with potentially many local optima. It is easy to end up in a globally sub-optimal region when running gradient-based optimization. We thus run maximization of the marginal loglikelihood multiple times, initializing from different hyperparameter values each time. This approach is adopted to increase the chance of ending up in a global maximum.

Specifically, to select the hyperparameters for each QS region, we run the following procedure for a given QS region of a sample of a dimension of a given patient:
\begin{enumerate}
    \item Select a hyperparameter initialisation grid $\mathcal{G} = \mathcal{S} \times \mathcal{P} \times \mathcal{L}_p \times \mathcal{L}_m$, with $\mathcal{S},\mathcal{P},\mathcal{L}_p$ and $\mathcal{L}_m$ being sets of potential values for $\sigma_f^2, p, \ell_p$ and $\ell_m$, respectively and set $\sigma^2$ to be 0.1 times the variance of the QS region (see below for details).
    \item For each $\theta_i \in \mathcal{G}$ initialize the hyperparameter vector to $\theta_i$ and maximize the marginal loglikelihood using gradient ascent on a grid constructed of 20\% randomly chosen points in the QS region. Using 20\% of the points prevents collinearity causing non-invertibility, and selecting them randomly encourages the parameters to better capture the dynamics of the full data of the QS region. We use 150 epochs with a learning rate of 0.1 and a Cholesky jitter of $0.001$.
    \item Take the fitted parameter vectors $\{\hat{\theta}_i\}$ and, for each $i$, produce 2500 samples $\{\Tilde{y}_i\}$ on a $[0,10]$ divided into 2560 points from a Gaussian process with kernel function \eqref{app_eqn:kernel} and hyperparameter $\hat{\theta}_i$. For each $\Tilde{y}_i$ compute the difference between the alpha, beta and theta power and also compute Kolomogorov-Smirnov (KS) statistic distance (1 minus the KS statistic) between the sample and the original QS region. Average these distances and select the hyperparameter index $i^*$ that gives the smallest distance. $\hat{\theta}_{i^*}$ is the hyperparameter for this region.
\end{enumerate}

\paragraph{Constructing $\mathcal{G}$.} 
A useful starting point for $\ell_m$ and $\ell_p$ can be found by examining the statistics of the training data. Very large length scales (e.g. $>10$) are often practically equivalent in the sense that they all practically describe a linear model. As the length scale governs how far two samples need to be separated in a Euclidean sense for their function values to have low correlation, a sensible heuristic is to examine the variance of the data. A sensible choice is to consider values between $\sqrt{\mathrm{var}(x)}/10$ and $\sqrt{\mathrm{var}(x)}$.

The noise variance and signal variance combine to explain the total variance of the sample. It is thus reasonable that $\sigma_f^2 + \sigma^2$ is approximately the variance of the sample. In cases where we have no prior information, it is often reasonable to set the noise variance to be smaller than the signal variance. After experimentation, we set $\sigma^2$ to be 0.1 times the variance of the region and $\sigma_f^2$ to be 0.9 times the same quantity.

Writing $\sigma_x = \sqrt{\mathrm{var}(x)}$, the result of these heuristics is that, when considering data $x$, that we run the above procedure on a grid $\mathcal{G} = \mathcal{S} \times \mathcal{P} \times \mathcal{L}_p \times \mathcal{L}_m$ constructed as
\[
\mathcal{G} = [0.9\sigma_x] \times [1,4,8] \times [1,8,16] \times [\sigma_x/0.1,\sigma_x/1,\sigma_x/10].
\]

\subsection{GP regression in \textit{Python}}

We use the GPyTorch library version 1.13 \cite{gardner2018gpytorch} for all experiments.

\section{Changepoint detection} \label{app_sec:CP_detection}

In this section, we detail the algorithm used to detect changepoints in the projection dimensions of each sample. Dropping subscripts of $n$ and superscripts of $(s)$ for clarity, consider a temporal score matrix $U \in \mathbb{R}^{T \times d}$. For each latent dimension $r\in\{1,\dots,d\}$, we consider the one-dimensional temporal score sequence
\[
U_r = (U_{1r},\dots,U_{Tr})^\top \in \mathbb{R}^T.
\]
Our aim is to segment each $U_r$ into contiguous \emph{quasi-stationary} intervals, i.e., regions within which a stationary time-series model is a reasonable approximation, while allowing changes in distributional properties across segments. We perform offline segmentation using two complementary stationarity/unit-root tests: the Kwiatkowski--Phillips--Schmidt--Shin (KPSS) test and the augmented Dickey--Fuller (ADF) test, both implemented via the \texttt{ct} procedure in the \texttt{statsmodel} package.

\paragraph{Overview of the segmentation principle.}
Both KPSS and ADF return evidence about stationarity on a candidate interval $[a,b]$ with $a,b \in \{0,\dots,T\}$ and $a < b$. We treat an interval as \emph{locally stationary} if it passes a stationarity check based on these tests (defined below), and we search for changepoints as locations where this local stationarity criterion fails. We set $a=0, b=256$ and increment each by 128 to check for stationarity using a sliding window approach. Concretely, for each dimension $r$ we construct changepoints
\[
0=\tau_{r,0} < \tau_{r,1} < \cdots < \tau_{r, K_r} < \tau_{r, K_r+1}=T.
\]

\paragraph{KPSS test (null: stationarity)}
The KPSS test evaluates the null hypothesis that a series is (level- or trend-) stationary against the alternative of a unit root (difference stationarity). On a candidate segment $\{U_{tr}\}_{t=a}^b$ of length $n=b-a+1$, KPSS is typically formulated via the decomposition
\[
U_{tr} = \mu_t + \varepsilon_t,
\]
where $\varepsilon_t$ is a stationary error and $\mu_t$ is either a constant (level-stationarity) or a deterministic trend (trend-stationarity) under the null. Here we considered trend-stationary as the null using the \texttt{statsmodel} package in \textit{Python}. Operationally, one fits the chosen deterministic component and obtains residuals $\hat{\varepsilon}_t$ on the segment. Let
\[
S_j = \sum_{t=a}^{a+j-1} \hat{\varepsilon}_t,\qquad j=1,\dots,n,
\]
denote cumulative residual sums. The KPSS statistic is
\begin{equation}\label{eq:kpss_stat}
\mathrm{KPSS} \;=\; \frac{1}{n^2\,\hat{\sigma}^2}\sum_{j=1}^{n} S_j^2,
\end{equation}
where $\hat{\sigma}^2$ is an estimator of the long-run variance of $\hat{\varepsilon}_t$. Large values of \eqref{eq:kpss_stat} provide evidence against the null of stationarity. In our segmentation logic, a segment \emph{passes} KPSS if it \emph{fails to reject} stationarity at the chosen significance level (here 5\%).

\paragraph{ADF test (null: unit root)}
The ADF test evaluates the null hypothesis that a series has a unit root (is non-stationary) against the alternative of stationarity. On a segment $\{U_{tr}\}_{t=a}^b$ (length $n$), ADF is typically based on the regression
\begin{equation}\label{eq:adf_regression}
\Delta U_{tr} = \alpha + \beta\,t + \gamma\,U_{t-1,r} + \sum_{i=1}^{p}\phi_i\,\Delta U_{t-i,r} + e_t,
\qquad t=a+1,\dots,b,
\end{equation}
where $\Delta U_{tr}=U_{tr}-U_{t-1,r}$, the optional $(\alpha,\beta t)$ terms allow for a constant and/or deterministic trend, and the lag order $p$ controls residual autocorrelation. The null hypothesis is $\gamma=0$ (unit root), and the alternative is $\gamma<0$ (stationary). The ADF test statistic is the $t$-statistic for $\hat{\gamma}$ in \eqref{eq:adf_regression}, compared against non-standard critical values. In our segmentation logic, a segment \emph{passes} ADF if it \emph{rejects} the unit-root null at the chosen significance level (here 5\%).

\paragraph{Combining KPSS and ADF for quasi-stationary segmentation.}
KPSS and ADF are complementary: KPSS tests \emph{stationarity as the null}, while ADF tests \emph{unit root as the null}. For each latent dimension $r$, we run the \texttt{ct} procedure separately with KPSS and with ADF, yielding two changepoint sets, denoted
\[
\{\tau^{\mathrm{KPSS}}_{r,j}\}_{j=1}^{K_r^{\mathrm{KPSS}}}
\qquad\text{and}\qquad
\{\tau^{\mathrm{ADF}}_{r,j}\}_{j=1}^{K_r^{\mathrm{ADF}}}.
\]
We then merge these sets to obtain a single set of changepoints. Specifically, whenever a KPSS changepoint and an ADF changepoint occur within 256 time points of one another, we treat them as referring to the same underlying transition and replace the pair by their midpoint (rounded to the nearest integer):
\[
\text{if } \big|\tau^{\mathrm{KPSS}}_{r,i}-\tau^{\mathrm{ADF}}_{r,j}\big|\le 256,
\quad\text{set}\quad
\tau_{r,\ell} \;=\; \left\lfloor \frac{\tau^{\mathrm{KPSS}}_{r,i}+\tau^{\mathrm{ADF}}_{r,j}}{2}\right\rceil .
\]
All remaining (unpaired) changepoints from either test are retained unchanged, and the resulting merged set is sorted to form
\[
0=\tau_{r,0} < \tau_{r,1} < \cdots < \tau_{r, K_r} < \tau_{r, K_r+1}=T.
\]
This yields a conservative consensus segmentation: changepoints supported by both tests are consolidated, while changepoints detected by only one test are still permitted. Because of computational issues with fitting a GP to a segment of more than 2560 samples, if no changepoint is observed for 10 s, we insert a CP manually at that point. This was not found to be required more than a small number of times.

\begin{figure}[h!]
    \centering
    \includegraphics[width=\textwidth]{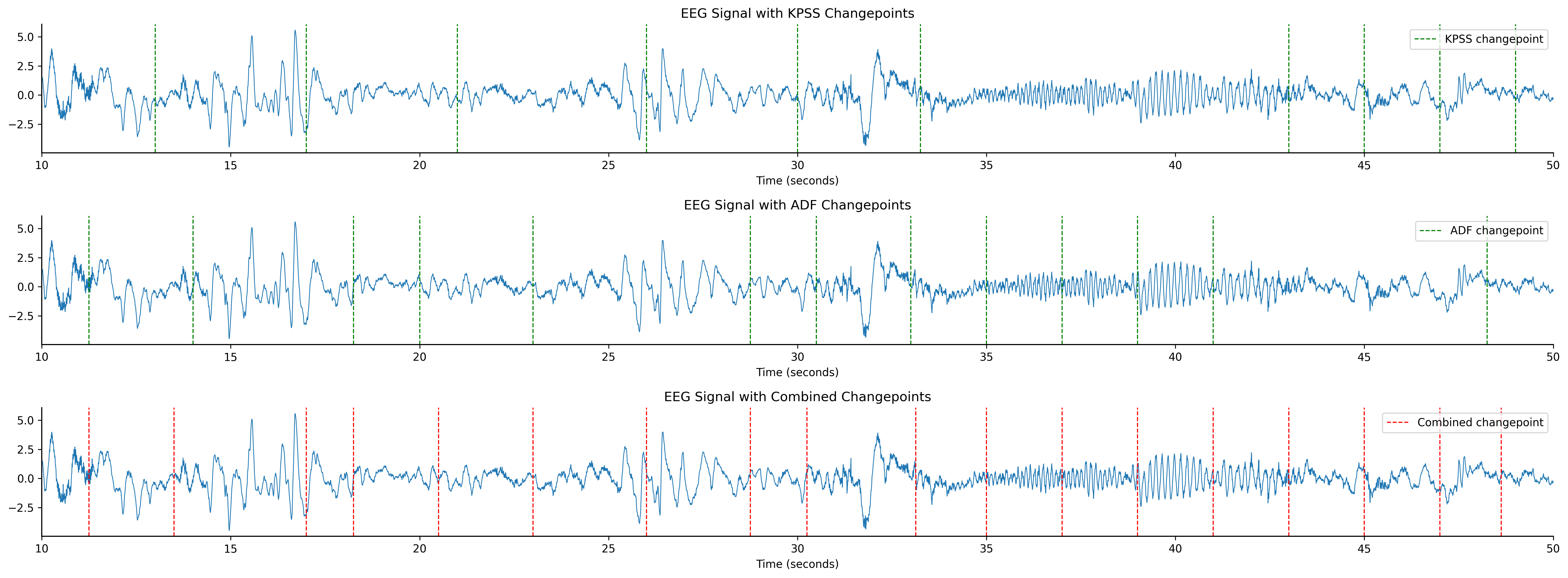}
    \caption{A channel of an EEG signal with changepoints plotted. In the first row are changepoints computed using KPSS, in the second using ADF, and in the third both KPSS and ADF.}
    \label{fig:cp_detection}
\end{figure}

This final set of changepoints partitions each latent score dimension $U_r$ into quasi-stationary regimes, which are subsequently modeled separately in later stages of the pipeline. In Figure~\ref{fig:cp_detection}, we show examples of changepoints computed using KPSS, ADF, and the merged set.

\section{Discretization of the hyperparameter space} \label{app_sec:discretisation}

\begin{figure}[h!]
    \centering
    \includegraphics[width=0.9\textwidth]{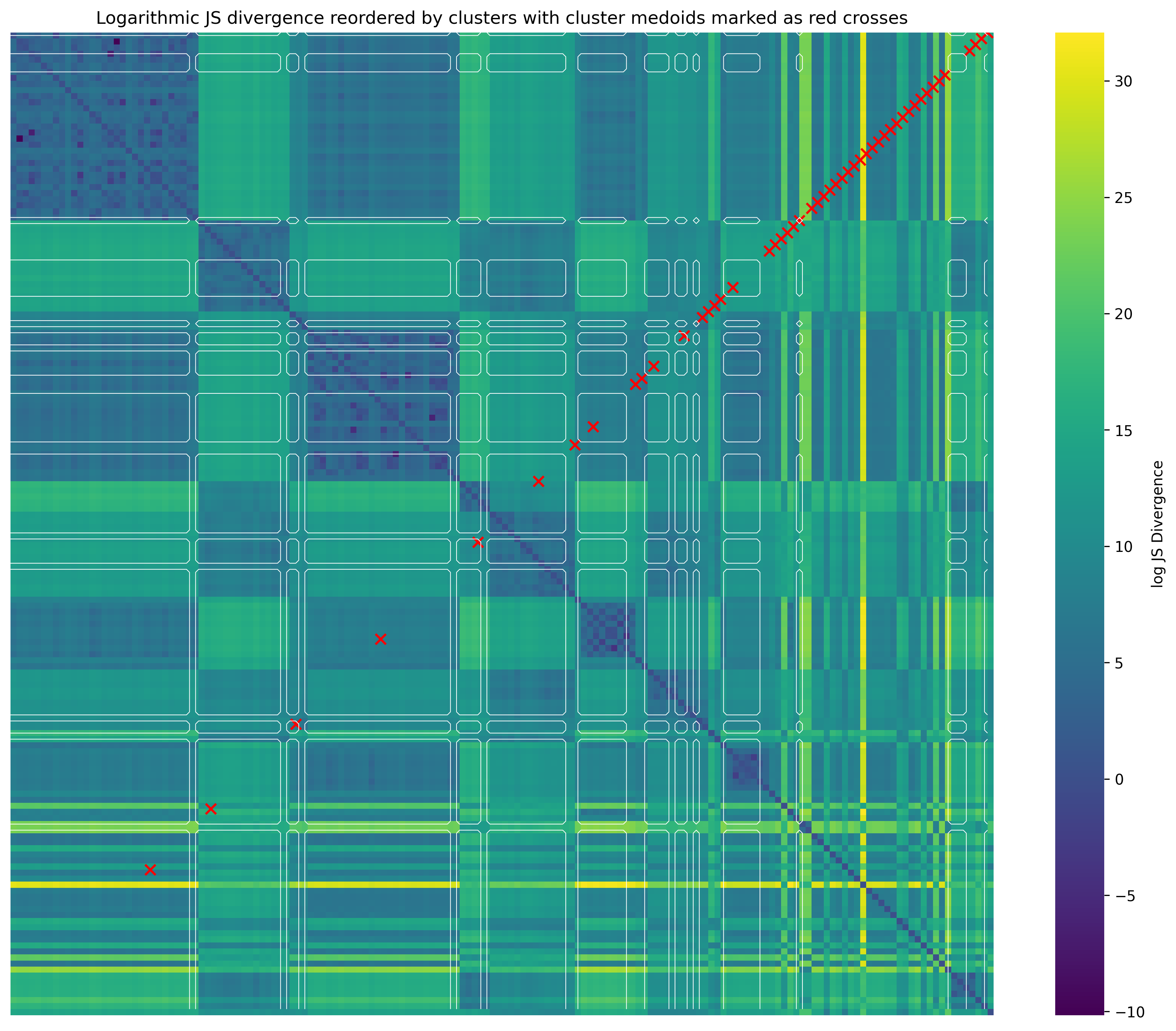}
    \caption{Log-transformed JS-divergence between Gaussians formed from all $J_n \choose 2$ pairs of hyperparameters, as described in Appendix \ref{app_sec:discretisation}.}
    \label{fig:JS_matrix_medoid}
\end{figure}

To discretize the hyperparameter space, we take all $J_n$ parameters for a patient and for each of the $J_n \choose 2$ pairs we parameterize the kernel function in \eqref{app_eqn:kernel} and form a 2560-dimensional multivariate Gaussian by evaluating \eqref{app_eqn:kernel} on a grid of 2560 points on $[0,10]$ and zero mean, as detailed in the main body in Stage 2. Figure \ref{fig:JS_matrix_medoid} shows the (log-transformed) JS-divergence between all pairwise Gaussian distributions. This is used as a similarity matrix to run agglomerative clustering using the average linkage. We use the \texttt{scikit-learn} implementation of this algorithm with $50$ clusters. Once the clusters are obtained, we compute the medoids of each cluster. These form our representative points.

\begin{figure}[h!]
    \centering
    \includegraphics[width=\textwidth]{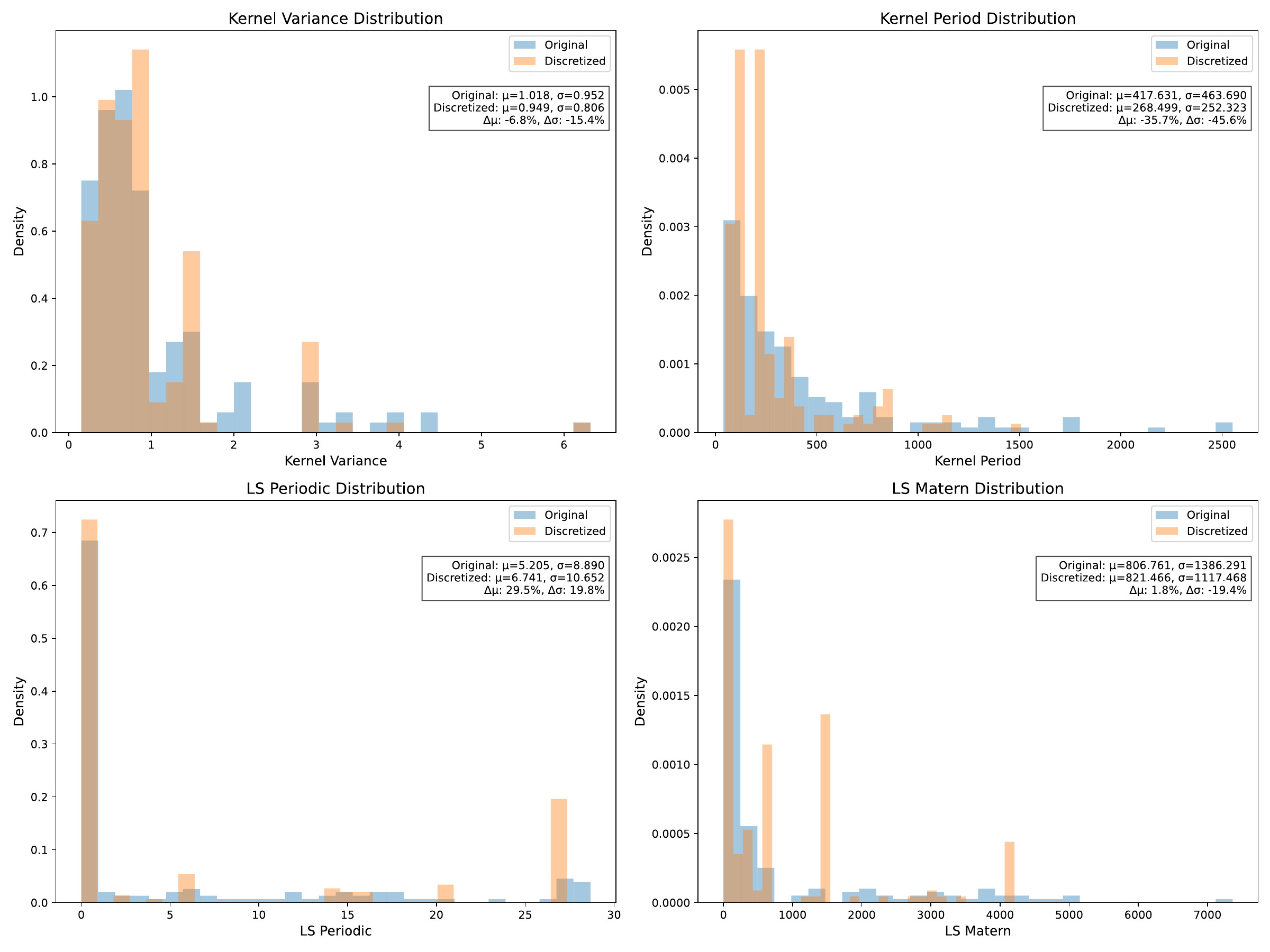}
    \caption{Histogram plots of the distributions of the 4 parameters of the kernel in \eqref{app_eqn:kernel}. In blue is the parameters in their continuous form, and in orange is once they are mapped to the medoid of their cluster.}
    \label{fig:discrete_param_scape}
\end{figure}

Figure \ref{fig:discrete_param_scape} shows the distributions of the sampled hyperparameters. In blue, we see the parameters in their continuous form, and in orange the parameters once they have been mapped to the medoid of their cluster, as described above. We see that the discretizing captures most of the behavior of the continuous setting, but removes some of the extreme values.


\section{Training details for the Conv-LSTM VAE domain adaptation network}
\label{app:vae_training}

This section describes the training procedure for the Conv-LSTM VAE used in Stage 3 of the proposed pipeline. The purpose of the Conv-LSTM VAE is to refine GP-generated synthetic EEG segments to better match the spectral structure and waveform of real EEG, while preserving the long-range temporal dependencies imposed by the preceding stages.

\paragraph{Training dataset construction.}

The Conv-LSTM VAE is trained using the one-to-one mapping between GP-surrogates and real EEG segments, obtained after Stage 2. That is, for each seizure recording, (i) we obtain the temporal scores via patient-specific SVD decomposition, (ii) sample temporal scores from the fitted GP posteriors conditioned on the empirically detected changepoints and learned kernel hyperparameters (without discretizing the kernel space or sampling new changepoint locations), and (iii) project the temporal scores back to the original signal space using the patient-specific spatial loadings. This procedure yields a surrogate GP-EEG $
\tilde{x}_{n,\mathrm{raw}}^{(s)}$  segment that `matches' the real segment $\tilde{x}_{n}^{(s)}$. 

The Conv-LSTM VAE is trained on the set of z-score normalized (normalization is done at the sample level, and not per channel) paired samples $(\tilde{x}_{n,\mathrm{raw}}^{(s)},  \tilde{x}_{n}^{(s)})$. Because the surrogate input is generated by sampling from a GP conditioned on the real signal's latent structure, the mapping preserves temporal alignment at the regime level while remaining stochastic at the waveform level, precluding trivial identity mappings.

\paragraph{Model architecture.}
The Conv-LSTM VAE consists of a shared convolutional encoder followed by a global latent bottleneck and channel-specific LSTM decoders. The encoder comprises three one-dimensional convolutional layers with kernel sizes $\{15,11,7\}$, strides of 2, batch normalization, and LeakyReLU activations. The latent distribution is parameterized by channel-wise convolutional layers producing mean and log-variance estimates, followed by global average pooling to obtain a single latent vector $\mathbf{z} \in \mathbb{R}^{256}$ per segment. 

Decoding is performed using $C=18$ independent LSTM-based decoders, one per EEG channel, each conditioned on the same shared latent code. This choice was inspired by \cite{seyfi2022generating}, who use the same input noise to channel-independent LSTM GANs for multivariate data generation. Each decoder consists of a single-layer LSTM with hidden dimension 256 followed by a linear projection to the full temporal resolution. A residual connection combines the decoded output with the input surrogate segment using a learnable mixing coefficient $\lambda$, initialized to 0.3.

\paragraph{Normalization and loss function.}
Both surrogate inputs and real targets are normalized on a per-segment basis using a single mean and standard deviation computed across all channels and time points. The Conv-LSTM VAE is trained to minimize the standard $\beta$-VAE objective:
\begin{equation}
\mathcal{L}_{\text{VAE}} =
\|\hat{x} - x\|_2^2
+ \beta \, \mathrm{KL}\big(q_\phi(z \mid \tilde{x}) \,\|\, \mathcal{N}(0, I)\big),
\end{equation}
with $\beta = 10^{-3}$. The reconstruction loss is computed as mean squared error on normalized signals.

\paragraph{Training procedure.}
The Conv-LSTM VAE is trained in a patient-specific manner using the AdamW optimizer with learning rate $5 \times 10^{-4}$ and weight decay $10^{-4}$. Training is performed for 1000 epochs with batch size 8, and gradient clipping at norm 1.0.

\paragraph{Use at inference.}
During synthetic data generation, new EEG segments are produced by sampling changepoint sequences and GP kernel states using the learned transition matrix (Stage 6 of the pipeline), followed by GP sampling within each regime. These synthetic temporal scores are projected back to the signal space using the patient-specific spatial loadings, yielding surrogate EEG segments that are then normalized and passed through the Conv-LSTM VAE. In Figures \ref{fig:eegification_siena} and \ref{fig:eegification_CHB}, we see synthetic seizures before and after being passed through the Conv-LSTM VAE. We see that before ``EEGifying'' the sample, the synthetic seizures are very smooth and are visually unlike true EEG data. We see in the second column that the Conv-LSTM VAE adds roughness and periodicity to the sample that is reflective of EEG data.

\begin{figure}[h!]
    \centering
    \includegraphics[width=\textwidth]{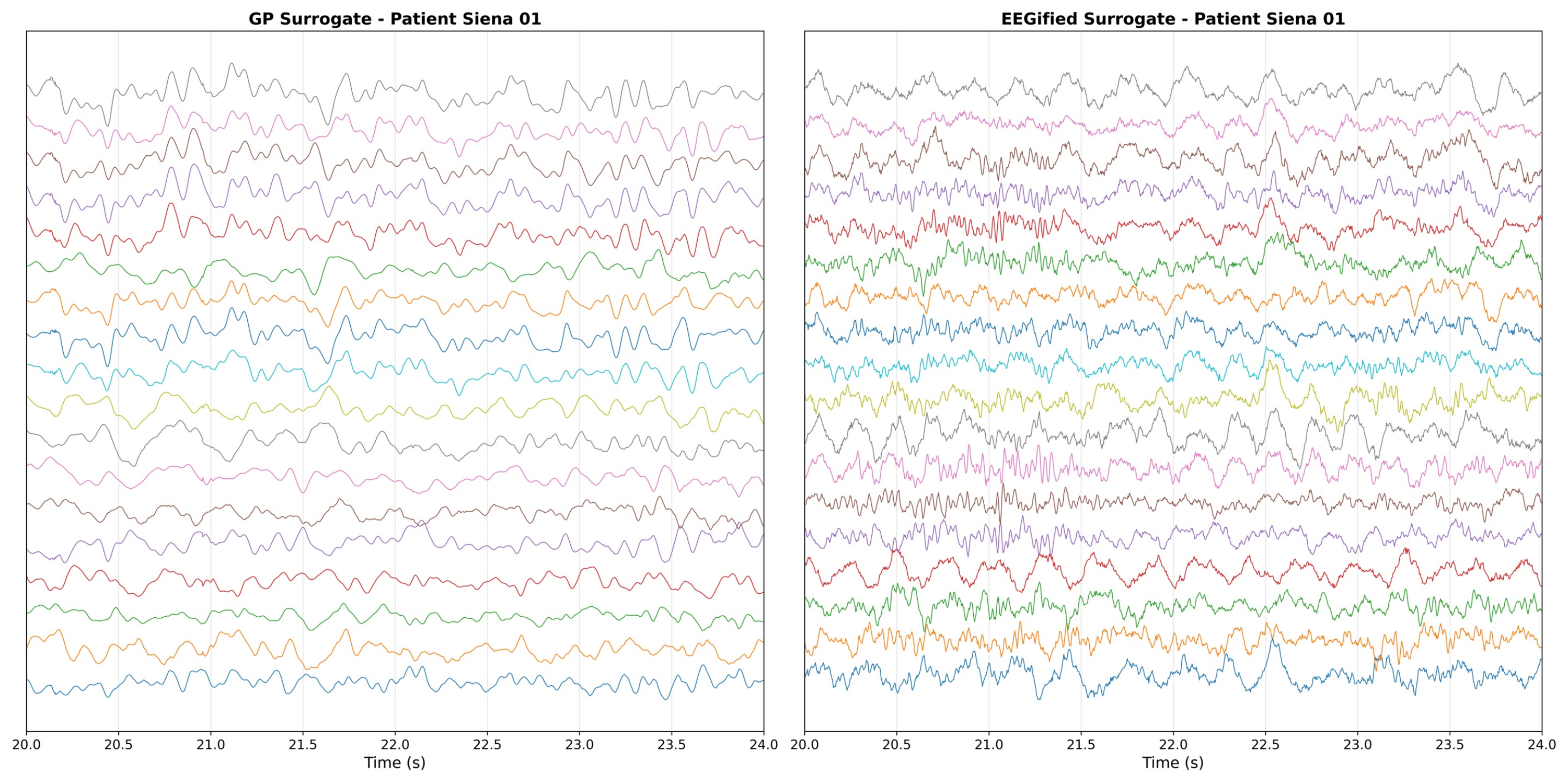}
    \includegraphics[width=\textwidth]{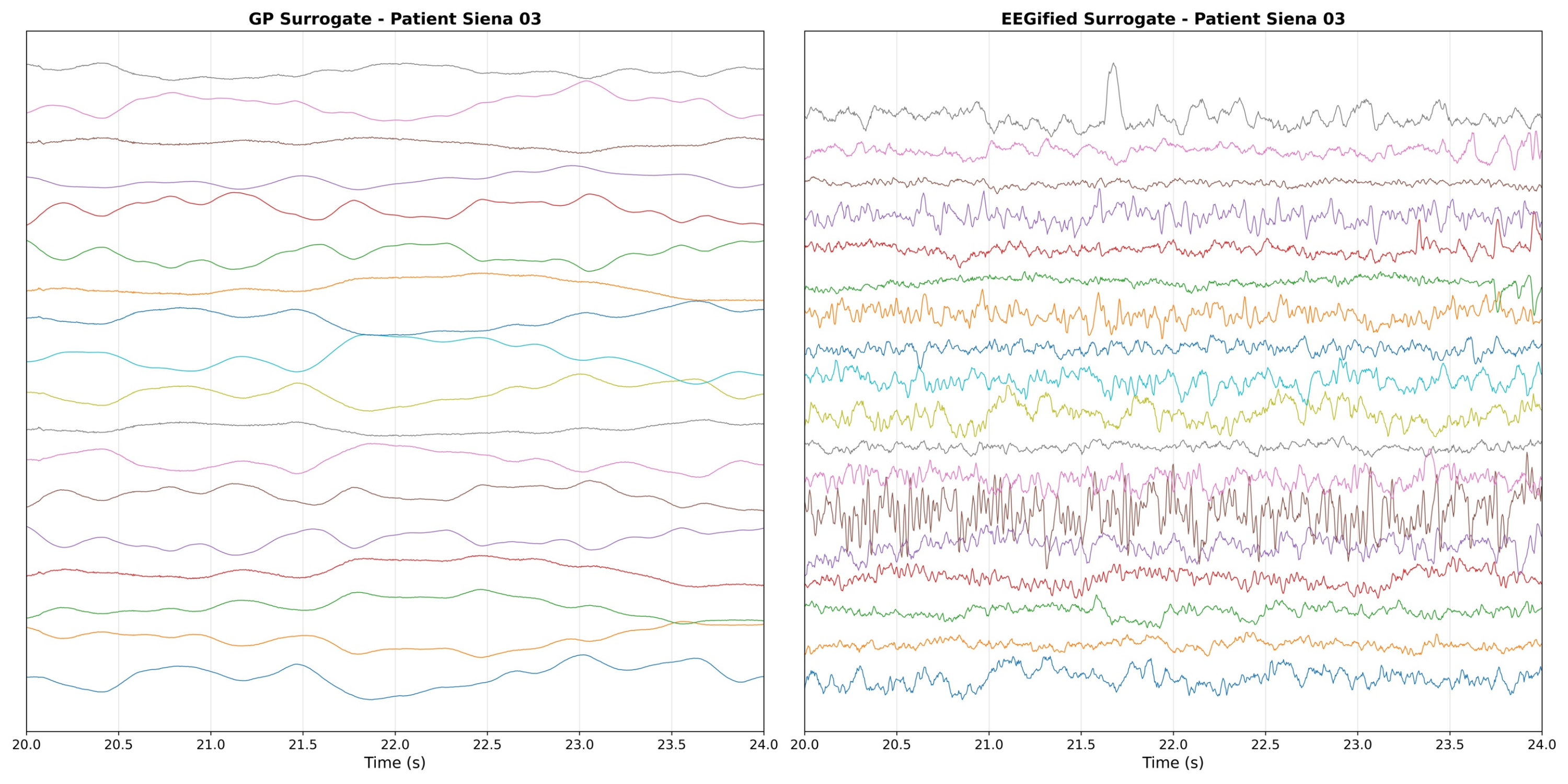}
    \caption{A comparison of the synthetic samples generated from a patient in the Siena dataset before and after being passed through the Conv-LSTM VAE. Each row is a sample, and the first and second column correspond to pre and post the Conv-LSTM VAE stage.}
    \label{fig:eegification_siena}
\end{figure}
\begin{figure}[h!]\ContinuedFloat
    \centering
    \includegraphics[width=\textwidth]{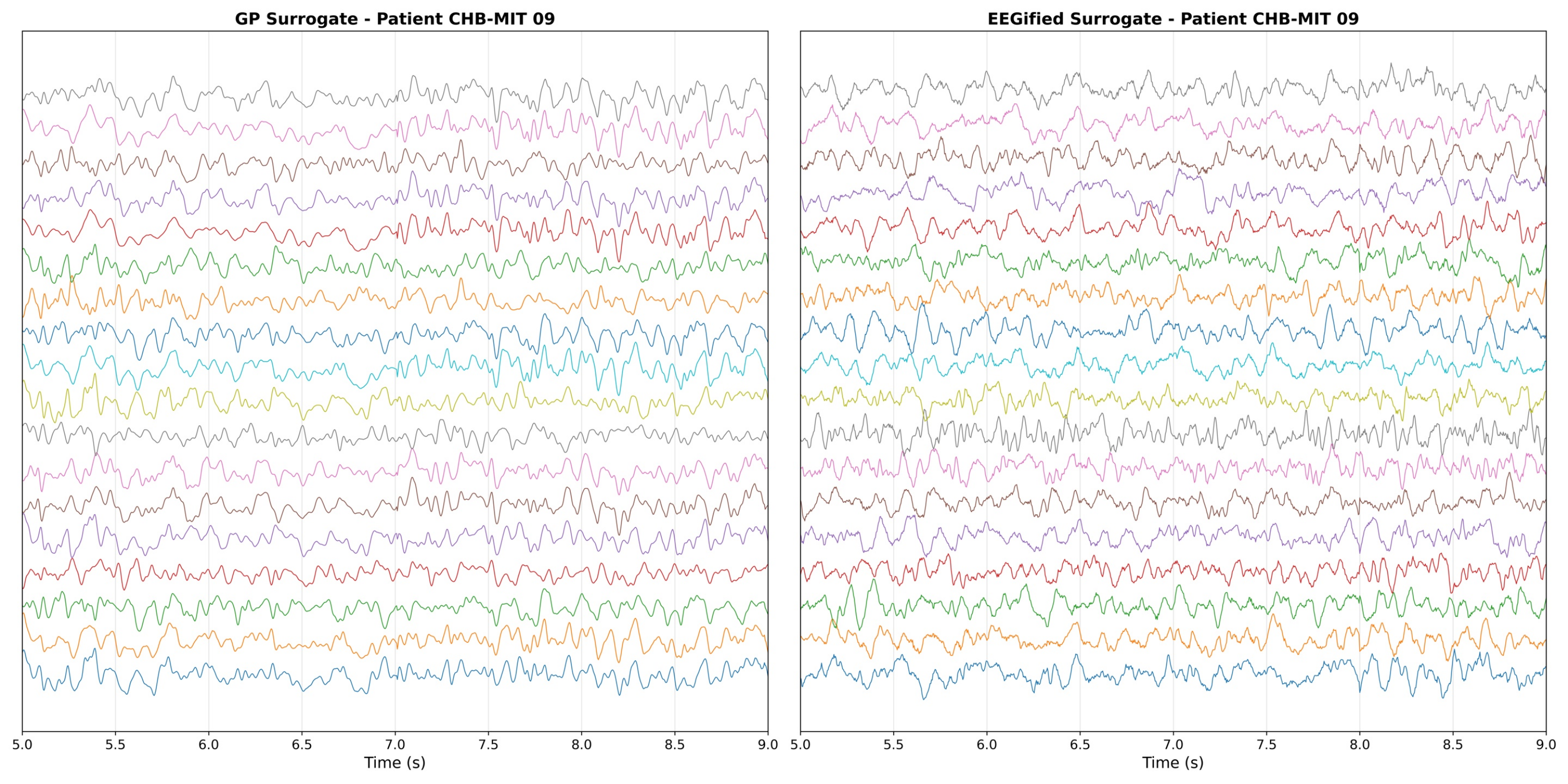}
    \includegraphics[width=\textwidth]{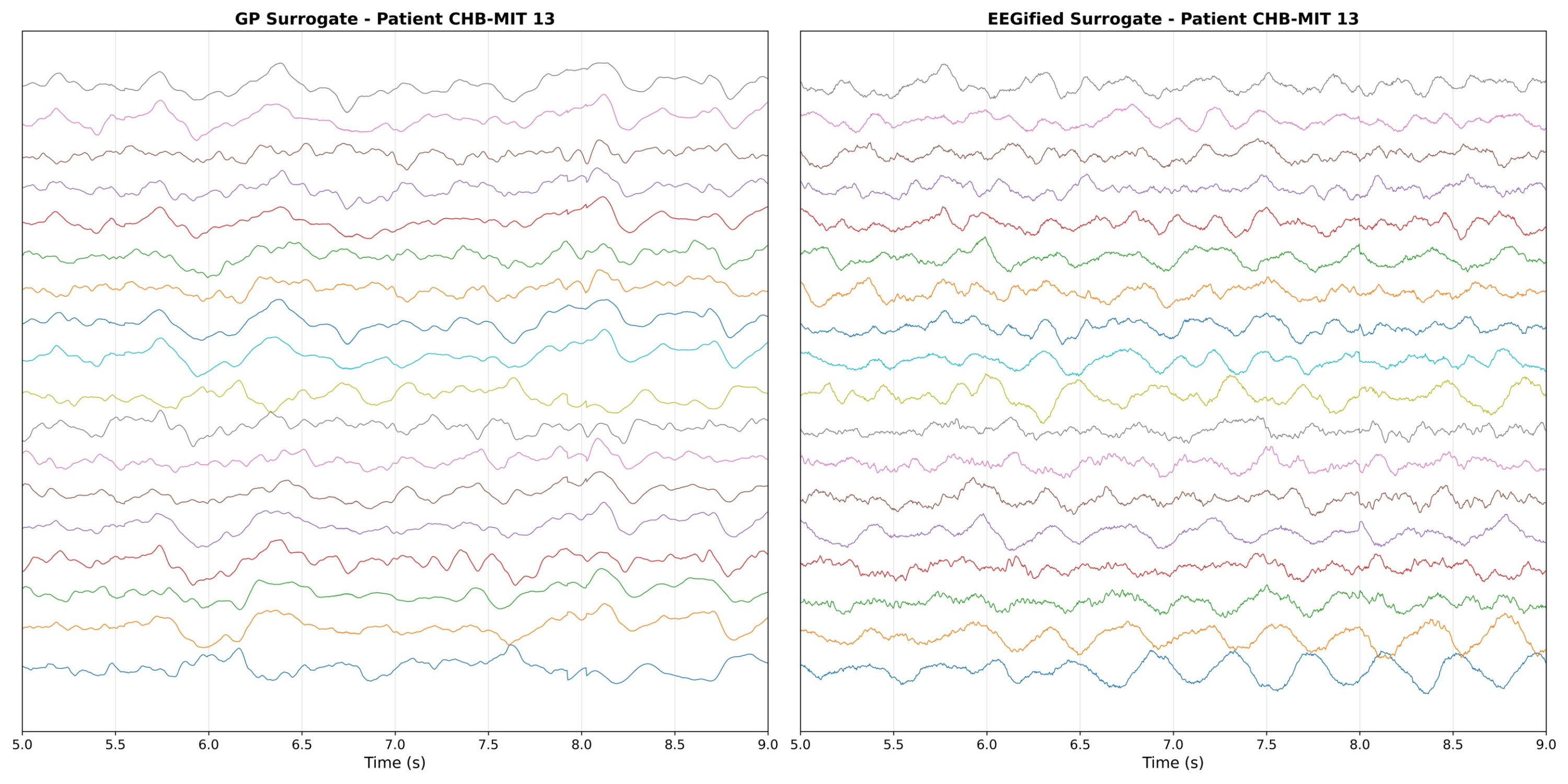}
    \caption{A comparison of the synthetic samples generated from a patient in the CHB-MIT dataset before and after being passed through the Conv-LSTM VAE. Each row is a sample, and the first and second column correspond to pre and post the Conv-LSTM VAE stage.}
    \label{fig:eegification_CHB}
\end{figure}

\section{Feature space analysis}\label{app_sec:feature_space}

Figures \ref{fig:feature_siena_1} and \ref{fig:feature_chbmit_1} depict the distribution of features of the real and synthetic EEG seizure segments across the Siena and CHB-MIT datasets respectively. Two complementary feature sets were extracted. The first is Catch22, a standardised collection of 22 canonical time-series features selected for their diversity, interpretability, and computational efficiency across a wide range of time-series data-mining tasks \cite{lubba2019catch22}. These features span a range of temporal and statistical properties including autocorrelation structure, distributional characteristics, entropy measures, and non-linear dynamics. The second feature set comprises EEG-specific features including relative band powers (delta, theta, alpha, and beta), the theta-alpha ratio, a slowing ratio quantifying the proportion of low-frequency activity, mean and maximum inter-channel correlation, line length, and spectral entropy. Together, these features capture both general time-series dynamics and characteristics specific to EEG. 

For each sample, features were computed independently for each channel and subsequently averaged across channels to yield a single value per feature. All samples were z-score normalized on a per-sample basis prior to feature extraction. 

\begin{table}[h]
\centering
\small
\caption{Catch22 feature names and corresponding descriptions, taken from \cite{lubba2019catch22}.}
\label{tab:catch22_feature_names}
\begin{tabularx}{\textwidth}{lX}
\toprule
\textbf{Catch22 feature name} & \textbf{Description} \\
\toprule
\multicolumn{2}{l}{\textbf{Distribution}} \\
DN\_HistogramMode\_5 & Mode of z-scored distribution (5-bin histogram) \\
DN\_HistogramMode\_10 & Mode of z-scored distribution (10-bin histogram) \\
\midrule
\multicolumn{2}{l}{\textbf{Simple temporal statistics}} \\
SB\_BinaryStats\_mean\_longstretch1 & Longest period of consecutive values above the mean \\
DN\_OutlierInclude\_p\_001\_mdrmd & Time intervals between successive extreme events above the mean \\
DN\_OutlierInclude\_n\_001\_mdrmd & Time intervals between successive extreme events below the mean \\
\midrule
\multicolumn{2}{l}{\textbf{Linear autocorrelation}} \\
CO\_f1ecac & First $1/e$ crossing of autocorrelation function \\
CO\_FirstMin\_ac & First minimum of autocorrelation function \\
SP\_Summaries\_welch\_rect\_area\_5\_1 & Total power in lowest fifth of frequencies in the Fourier power spectrum \\
SP\_Summaries\_welch\_rect\_centroid & Centroid of the Fourier power spectrum \\
FC\_LocalSimple\_mean3\_stderr & Mean error from a rolling 3-sample mean forecasting \\
\midrule
\multicolumn{2}{l}{\textbf{Nonlinear autocorrelation}} \\
CO\_trev\_1\_num & Time-reversibility statistic \\
CO\_HistogramAMI\_even\_2\_5 & Automutual information \\
IN\_AutoMutualInfoStats\_40\_gaussian\_fmmi & First minimum of the automutual information function \\
\midrule
\multicolumn{2}{l}{\textbf{Successive differences}} \\
MD\_hrv\_classic\_pnn40 & Proportion of successive differences exceeding threshold \\
SB\_BinaryStats\_diff\_longstretch0 & Longest period of successive incremental decreases \\
SB\_MotifThree\_quantile\_hh & Shannon entropy of two successive letters in equiprobable 3-letter symbolization \\
FC\_LocalSimple\_mean1\_tauresrat & Change in correlation length after iterative differencing \\
CO\_Embed2\_Dist\_tau\_d\_expfit\_meandiff & Exponential fit to successive distances in 2-d embedding space \\
\midrule
\multicolumn{2}{l}{\textbf{Fluctuation Analysis}} \\
SC\_FluctAnal\_2\_dfa\_50\_1\_2\_logi\_prop\_r1 & Proportion of slower timescale fluctuations that scale with DFA (50\% sampling) \\
SC\_FluctAnal\_2\_rsrangefit\_50\_1\_logi\_prop\_r1 & Proportion of slower timescale fluctuations that scale with linearly rescaled range fits \\
\midrule
\multicolumn{2}{l}{\textbf{Others}} \\
SB\_TransitionMatrix\_3ac\_sumdiagcov & Trace of covariance of transition matrix between symbols in 3-letter alphabet \\
PD\_PeriodicityWang\_th0\_01 & Periodicity measure \\
\bottomrule
\end{tabularx}
\end{table}

\begin{figure}[h!]
    \centering
    \includegraphics[width=\textwidth]{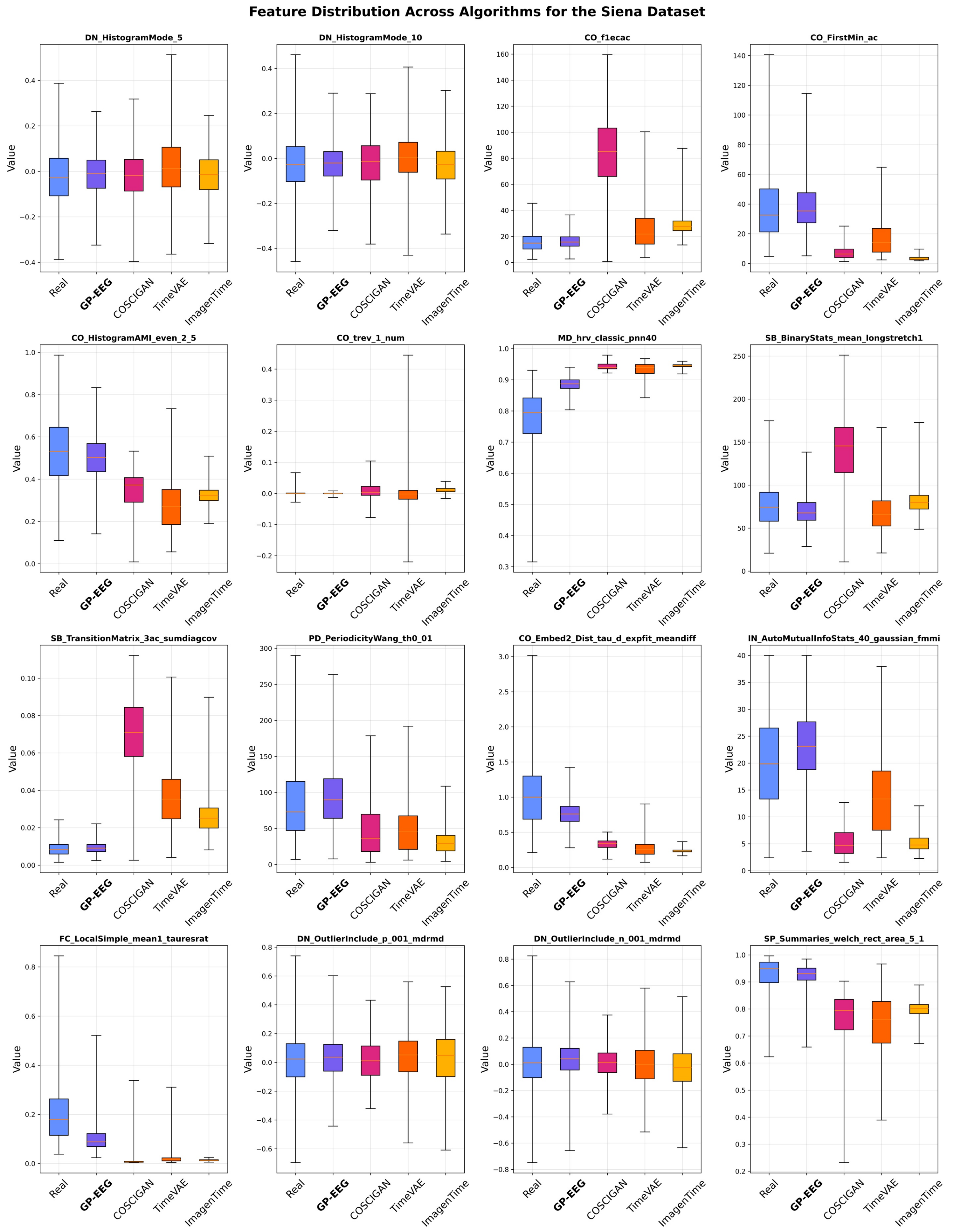}
    \caption{Distribution of Catch22 and EEG-specific features across real and synthetic seizure segments for the Siena dataset. Box plots show median (red line), interquartile range (box), and full range (whiskers). Detailed descriptions of the Catch22 features can be found in Table \ref{tab:catch22_feature_names}}
    \label{fig:feature_siena_1}
\end{figure}
\begin{figure}[h!]\ContinuedFloat
    \centering
    \includegraphics[width=\textwidth]{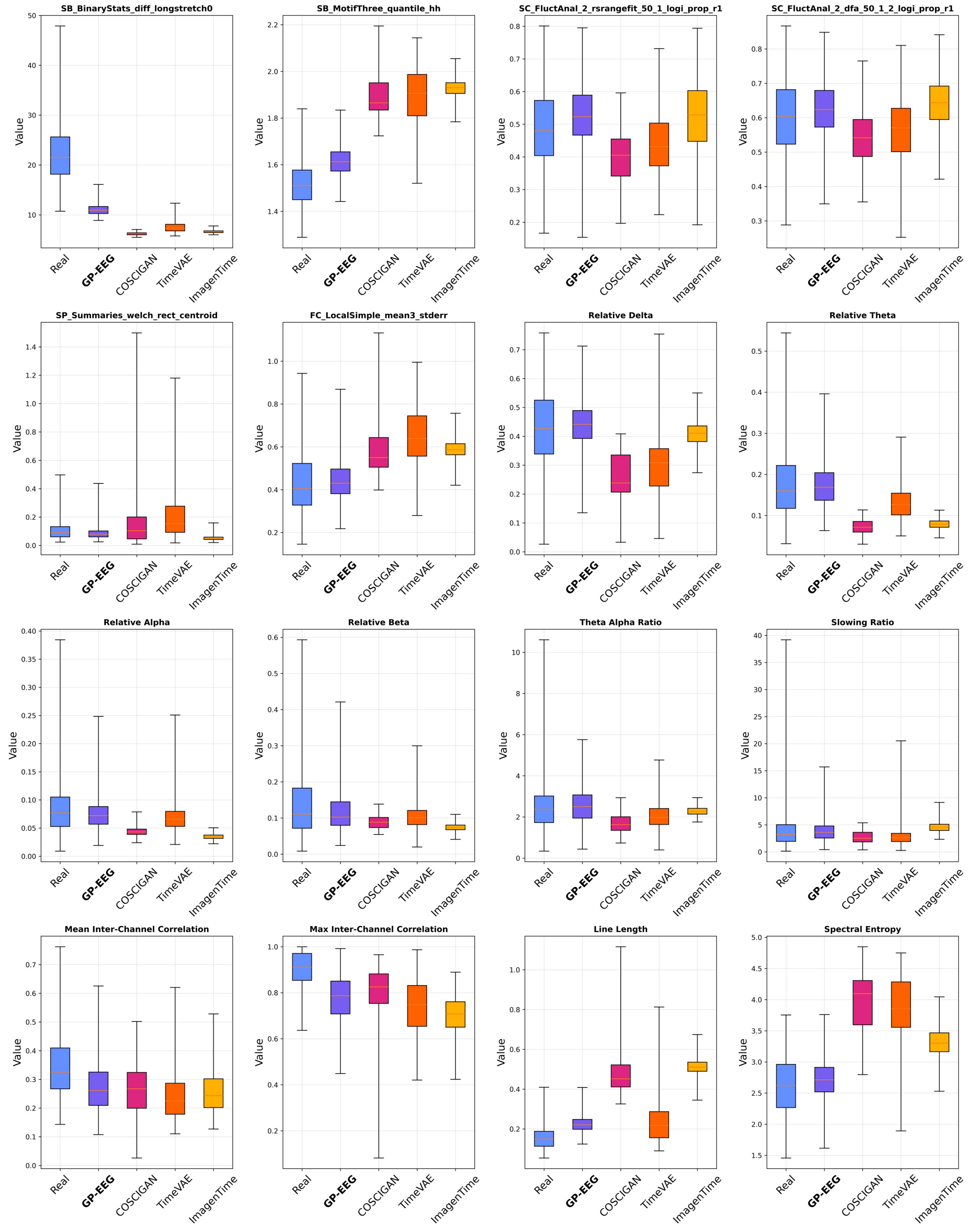}
    \caption{(continued) Distribution of Catch22 and EEG-specific features across real and synthetic seizure segments for the Siena dataset. Box plots show median (red line), interquartile range (box), and full range (whiskers). Detailed descriptions of the Catch22 features can be found in Table \ref{tab:catch22_feature_names}}
    \label{fig:feature_siena_2}
\end{figure}

\begin{figure}[h!]
    \centering
    \includegraphics[width=\textwidth]{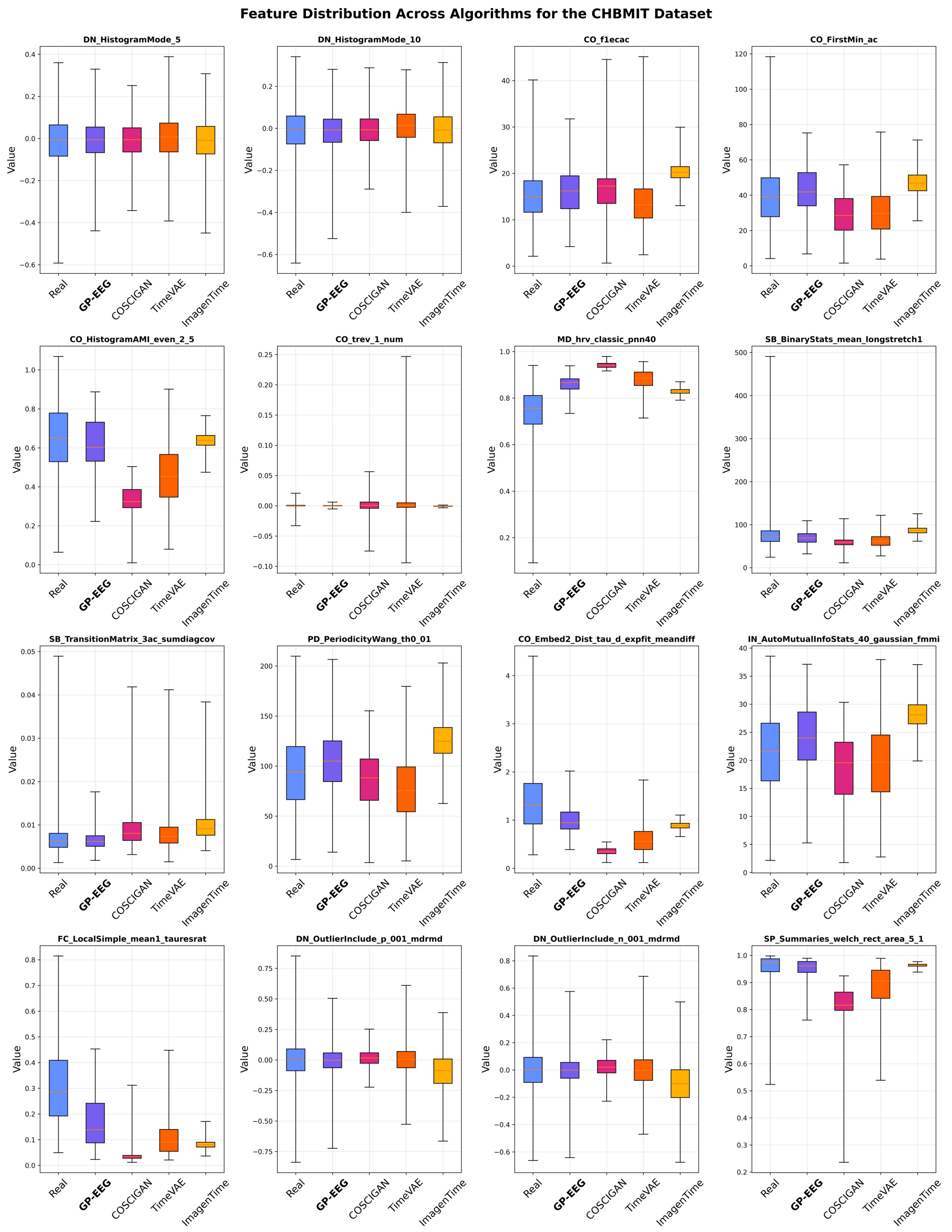}
    \caption{Distribution of Catch22 and EEG-specific features across real and synthetic seizure segments for the CHB-MIT dataset. Box plots show median (red line), interquartile range (box), and full range (whiskers). Detailed descriptions of the Catch22 features can be found in Table \ref{tab:catch22_feature_names}}
    \label{fig:feature_chbmit_1}
\end{figure}
\begin{figure}[h!]\ContinuedFloat
    \centering
    \includegraphics[width=\textwidth]{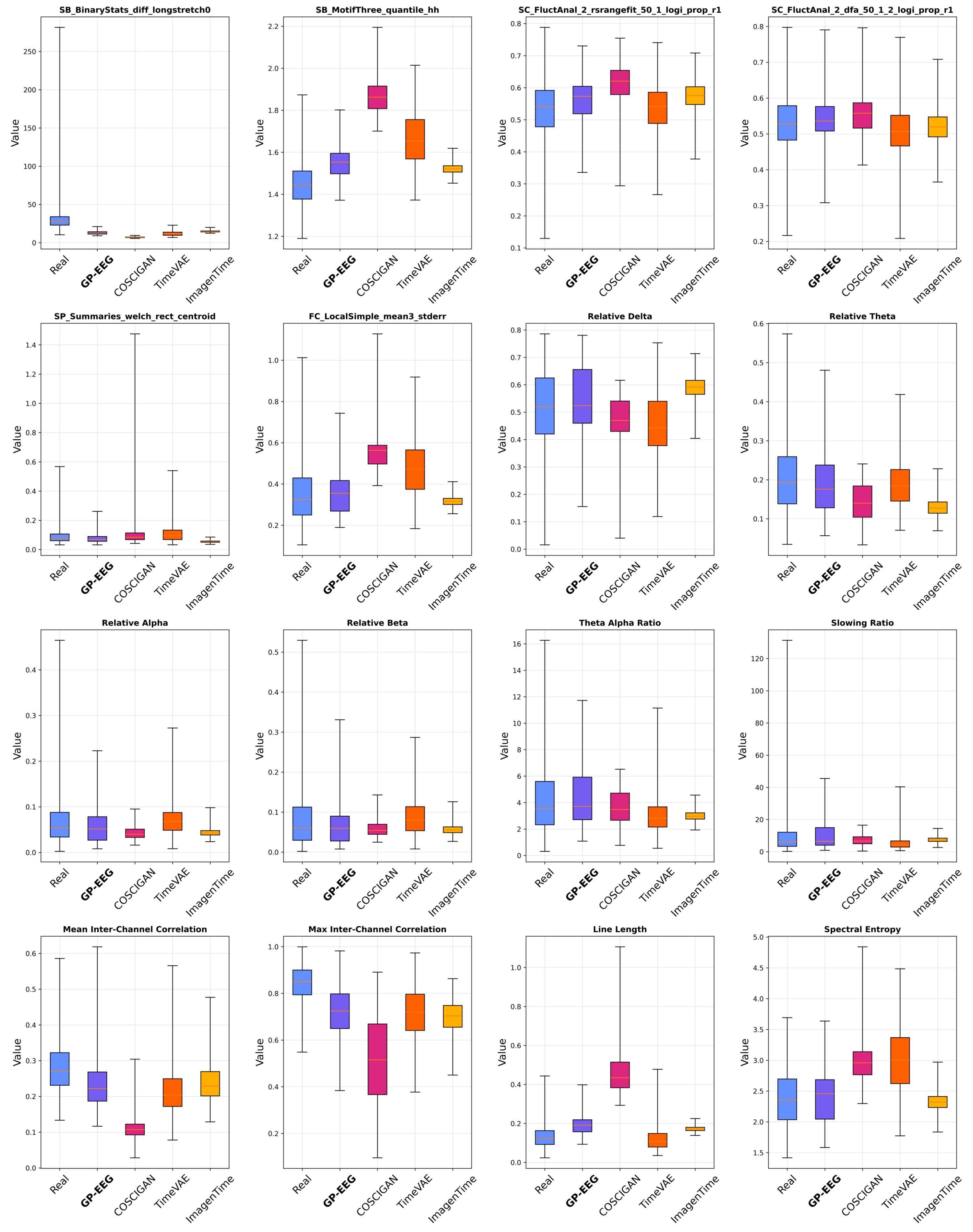}
    \caption{(continued) Distribution of Catch22 and EEG-specific features across real and synthetic seizure segments for the CHB-MIT dataset. Box plots show median (red line), interquartile range (box), and full range (whiskers). Detailed descriptions of the Catch22 features can be found in Table \ref{tab:catch22_feature_names}}
    \label{fig:feature_chbmit_2}
\end{figure}

\section{Baseline implementation details }\label{app_sec:baseline_hyperparams}

\paragraph{TimeVAE.}
We implement the base TimeVAE algorithm using the code provided by the author in this \href{https://github.com/abudesai/timeVAE}{GitHub}. Unless stated otherwise, all hyperparameters follow the default configuration.
TimeVAE is a convolutional variational autoencoder with residual connections and a Gaussian latent space of dimension $8$. The encoder and decoder consist of stacked 1D convolutional layers with channel sizes $\{50,100,200\}$. No trend component or custom seasonalities are used. Training minimizes the standard VAE objective with a reconstruction loss weight of $3.0$. Optimization is performed using Adam with its default learning rate of $10^{-3}$ and a batch size of $16$. Models are trained for up to $300$ epochs with early stopping (patience $50$, minimum improvement $10^{-2}$) and learning rate reduction on plateau (factor $0.5$, patience $30$). 

\paragraph{COSCI-GAN.}
We use the official COSCI-GAN implementation released by the authors in this \href{https://github.com/aliseyfi75/COSCI-GAN}{GitHub}. Both the generator and discriminator are implemented using LSTM-based architectures.
A central discriminator is enabled to explicitly capture cross-channel dependencies, using an MLP-based discriminator as the authors found MLP cross-channel discriminators to perform better than LSTM cross-channel discriminators. Models are trained for $300$ epochs with a batch size of $32$ on sequences of length $1024$ across $18$ channels.
The adversarial loss is binary cross-entropy, with a correlation regularization weight $\gamma = 5.0$, which was found to be optimal by the authors. Optimization uses Adam with learning rates $10^{-3}$ for the generator and discriminator, and $10^{-4}$ for the central discriminator as per the official implementation. The input noise dimension is set to $32$.

\paragraph{ImagenTime.}
We use the official ImagenTime implementation released in the GitHub \href{https://github.com/azencot-group/ImagenTime}{GitHub} in the unconditional generation setting. ImagenTime is a diffusion-based generative model that maps multivariate time series to images via a time--frequency representation. Input sequences of length $1024$ sampled at $256$\,Hz are transformed using STFT embeddings. We use $n_{\mathrm{fft}}=128$ and hop length $16$, corresponding to $0.5$\,s analysis windows with $93.75\%$ overlap, providing a balanced time--frequency resolution for non-stationary EEG signals. This yields $36$ input channels (real and imaginary components of $18$ EEG channels).
The denoising network is a U-Net with base channel width $128$, channel multipliers $\{1,2,4,4\}$, and attention at resolutions $\{32,16,8\}$. Diffusion is performed over $18$ steps, with exponential moving average (EMA) of model parameters enabled after a warmup of $100$ epochs. Models are trained for $1000$ epochs with a batch size of $16$ using AdamW (learning rate $10^{-4}$, weight decay $10^{-5}$).

\section{MDD metric}\label{app_sec:mdd_metric}

Here we describe how the MDD is computed. Throughout, we consider a patient $n$ and drop all subscripts or superscripts of $n$. Suppose we have $S$ samples $\{x^{(s)}\}_{s=1}^S$ and we generate $S$ synthetic samples $\{\Tilde{x}^{(s)}\}$ as described in the main body. Each sample and synthetic time series is defined on $\mathbb{R}^{T\times C}$, with $T$ indexing time and $C$ indexing channels. Let $(x^{(s)})_{t,c}$ and $(\tilde{x}^{(s)})_{t,c}$ denote the $(t,c)$-th entry of each matrix. Form two vectors $y^{t,c}, \tilde{y}^{t,c}\in \mathbb{R}^s$, which are formed by taking the values across the $S$ samples in channel $c$ and time point $t$. For a fixed number of bins $B$, form intervals $\{\mathcal{B}^{t,c}_{b}\}_{b=1}^B$, and compute the total distance as:
\[
    D = \frac{1}{TC}\sum_{t=1}^{T}\sum_{c=1}^C D_{t,c}, \quad \text{where} \quad D_{t,c} := \frac{1}{S}\sum_{b=1}^{B}\sum_{s=1}^{S}\left(\boldsymbol{1}\{y^{t,c}_s \in \mathcal{B}^{t,c}_b\} - \boldsymbol{1}\{\tilde{y}^{t,c}_s \in \mathcal{B}^{t,c}_b\}\right),
\]
where $y_{s}^{t,c}$ is the $s$-th entry of $y^{t,c}$, and analogous notation is used for $\tilde{y}^{t,c}$.

\section{Train-on-synthetic, test-on-real evaluation}
\label{app:tstr}

We assess the usefulness of the proposed synthetic EEG data using a train-on-synthetic, test-on-real (TSTR) evaluation protocol based on a seizure detection task.

\paragraph{Cross-validation protocol.}
Evaluation is performed using leave-one-patient-out (LOPO) cross-validation. In each fold, one patient is held out for testing, while data from all remaining patients are used for training. The test set always consists exclusively of real EEG recordings (seizure and background) from the held-out patient.

\paragraph{Training data.}
For training, we compare two conditions: (i) models trained on real seizure EEG segments and real background EEG segments from the training patients, and (ii) models trained on synthetic seizure segments generated by the proposed framework together with real background EEG segments from the same patients. 

\paragraph{Classifier and training procedure.}
EEGNet4,2 \cite{lawhern2018eegnet} classifiers are trained using a kernel size of 512, a dropout of 0.3, the Adam optimizer with learning rate $10^{-4}$ and weight decay $10^{-3}$ and gradient clipping at norm 1.0. Training is performed for 50 epochs with batch size 64. Inputs are z-score standardized per-segments. Data was balanced for training and testing, and identical background segments were used in all experiments. 

\paragraph{Evaluation metrics.}
Performance is evaluated on the held-out real EEG test set using accuracy, precision, recall, F1-score, and area under the ROC curve (AUC). Reported results correspond to the mean and standard deviation across LOPO folds. Table \ref{app_tab:metrics_tstr} details the results of this experiment, while Table \ref{tab:metrics_tstr} in the main body provides the differences in performance.

\begin{table}[h]
\caption{Results of the baseline (TRTR) and TSTR experiments. }
\label{app_tab:metrics_tstr}
\begin{center}
\begin{small}
\begin{tabular}{lccccc}
\toprule
\multicolumn{6}{c}{\textbf{Siena}} \\
\midrule
Method & Acc (\%) & Pre (\%) & Rec (\%) & F1 (\%) & AUC (\%)\\
\midrule
Baseline & 74.51 & 77.20 & 71.31 & 71.64 & 86.37 \\
COSCI-GAN     & 49.44 & 10.71 & 0.33 & 0.63 & 50.66 \\
TimeVAE      & 50.00 & 0.00 & 0.00 & 0.00 & 55.15 \\
ImagenTime    & 50.74 & 35.71 & 1.70 & 3.22 & 55.42\\
GP-EEG (ours) & 74.83 & 83.66 & 62.86 & 68.59 & 84.86\\
\midrule
\multicolumn{6}{c}{\textbf{CHB-MIT}} \\
\midrule
Method & Acc (\%) & Pre (\%) & Rec (\%) & F1 (\%) & AUC (\%) \\
\midrule
Baseline & 71.69 & 74.23 & 64.57 & 67.54 & 78.55 \\
COSCI-GAN     & 49.98 & 0.00 & 0.00 & 0.00 & 49.31\\
TimeVAE       & 50.03 & 4.55 & 0.07 & 0.13 & 52.59 \\
ImagenTime    & 50.01 & 4.55 & 0.07 & 0.13 & 49.87 \\
GP-EEG (ours) & 65.94 & 71.03 & 51.22 & 57.75 & 74.49 \\
\bottomrule
\end{tabular}
\end{small}
\end{center}
\end{table}

\section{Data augmentation evaluation}
\label{app:augmentation_eval}

We provide additional details for the data augmentation evaluation presented in Section \ref{sec:data_augmentation_eval}.

\paragraph{Cross-validation protocol.}
Evaluation is conducted using leave-one-patient-out (LOPO) cross-validation, as for the TSTR scheme.

\paragraph{Training data.}
For training, we compare two conditions: (i) models trained on real seizure EEG segments and real background EEG segments from the training patients, and (ii) models trained on synthetic seizure EEG segments generated by the proposed framework together with real background EEG segments from the same patients, and real seizure EEG segments. Synthetic data is used only during training and is never included in the test set. Data is z-score normalized per sample (not per channel).

\paragraph{Classifier and training procedure.}
EEGNet4,2 \cite{lawhern2018eegnet} classifiers are trained using a kernel size of 512, a dropout of 0.3, the Adam optimizer with learning rate $10^{-4}$ and weight decay $10^{-4}$. Training is performed for 50 epochs with batch size 64. Inputs are z-score normalized per segment. To address class imbalance, class-weighted cross-entropy loss is employed, with weights computed from the training data distribution in each fold. As for the TSTR scheme, identical background segments are used for each patients across all experiments. 

\paragraph{Evaluation metrics.}
Performance is evaluated on the held-out real EEG test set using accuracy, precision, recall, F1-score, and area under the ROC curve (AUC). Reported results correspond to the mean and standard deviation across LOPO folds. Table \ref{app_tab:metrics_augmentation} details the results of this experiment, while Table \ref{tab:metrics_augmentation} in the main body provides the differences in performance between the augmented and baseline runs.

\begin{table}[h]
\caption{Results of the baseline and augmented experiments.}
\label{app_tab:metrics_augmentation}
\begin{center}
\begin{small}
\begin{tabular}{lccccc}
\toprule
\multicolumn{6}{c}{\textbf{Siena}} \\
\midrule
Method & Acc (\%) & Pre (\%) & Rec (\%) & F1 (\%) & AUC (\%)\\
\midrule
Baseline &  74.35 & 77.16 & 70.98 & 71.44 & 86.36 \\
COSCI-GAN     & 77.30 & 87.91 & 63.98 & 70.90 & 89.37 \\
TimeVAE      & 71.27 & 77.28 & 51.54 & 58.73 & 85.71 \\
ImagenTime   & 76.04 & 83.45 & 65.51 & 71.17 & 86.46 \\
GP-EEG (ours) & 79.28 & 85.86 & 72.24 & 76.55 & 89.15 \\
\midrule
\multicolumn{6}{c}{\textbf{CHB-MIT}} \\
\midrule
Method & Acc (\%) & Pre (\%) & Rec (\%) & F1 (\%) & AUC (\%) \\
\midrule
Baseline & 71.84 &  74.57 & 64.45 & 67.63 &  78.55\\
COSCI-GAN     & 68.84 & 81.15 & 47.22 & 57.11 & 78.89 \\
TimeVAE       & 67.85 & 84.74 &  42.13 & 53.41 & 78.51\\
ImagenTime    & 70.95 & 78.25 & 56.29 & 63.93 & 79.02 \\
GP-EEG (ours) & 73.60 & 76.94 & 66.71 & 70.38 & 80.52 \\
\bottomrule
\end{tabular}
\end{small}
\end{center}
\end{table}

\end{document}